\documentclass[aps,prb,showpacs,twocolumn,superscriptaddress]{revtex4-1}
\usepackage{amsmath,amssymb}
\usepackage{graphicx}
\usepackage{dcolumn}
\usepackage{bm}
\usepackage{dsfont}

\begin{document}
\title{Exact Dynamical Correlations of Hard-Core Anyons in One-Dimensional Lattices }
\author{Qing-Wei Wang}
\email[]{qingweiwang2012@163.com}
\affiliation{School of Information Engineering, Zhejiang Ocean University, Zhoushan, Zhejiang 316022, China}
\affiliation{Key Laboratory of Oceanographic Big Data Mining \& Application of Zhejiang Province,
Zhejiang Ocean University, Zhoushan, Zhejiang 316022, China}

\date{\today}
\begin{abstract}
  The dynamical correlations of a strongly correlated system is an essential ingredient to describe its non-equilibrium properties. We present a general method to calculate exactly the dynamical correlations of hard-core anyons in one-dimensional lattices, valid for any type of confining potential and any temperature. We obtain exact explicit expressions of the Green's function, the spectral function, and the out-of-time-ordered correlators (OTOCs).
  We find that the anyonic spectral function displays three main singularity lines which can be explained as a double spectrum in analogy to the Lieb-Liniger gas. The dispersion relations of these lines can be given explicitly and they cross at a \emph{hot point} $(q_m,\omega_m)$, which induces a peak in the momentum distribution function at $q_m$ and a power-law singularity in the local spectral function at $\omega_m$. We also find that the anyonic statistics can induces spatial asymmetry in the Green's function, its spectrum, and the OTOC. Moreover, the information spreading characterized by the OTOCs shows light-cone dynamics, asymmetric for general statistics and low temperatures, but symmetric at infinite temperature. Our results pave the way toward studying the non-equilibrium dynamics of hard-core anyons and experimentally probing anyonic statistics through spectral functions.
\end{abstract}

\maketitle
\section{Introduction}
Quantum particles can be classified as either bosons or fermions by their exchange statistics. However, Abelian anyons characterized by fractional statistics can also emerge in certain circumstances \cite{nuovo1977,JMP22-1664,PhysRevLett.48.1144,PhysRevLett.48.1559}, and play an important role in modern condensed-matter physics, such as fractional quantum Hall effect \cite{PhysRevLett.50.1395,PhysRevLett.52.1583,PhysRevLett.53.722}, topological quantum computing \cite{KITAEV2003,PhysRevLett.94.166802,RevModPhys.80.1083,science339-1179}, and  spin liquids \cite{KITAEV20062,PhysRevLett.99.247203}. Although originally proposed for two-dimensional systems, the concept of fractional statistics and anyons has been generalized to arbitrary dimensions \cite{PhysRevLett.67.937,PhysRevLett.66.1529}.
Especially, the physics of Abelian anyons in one dimension (1D) has recently attracted many theoretical interests \cite{PhysRevLett.73.1574,PhysRevLett.73.3331,PhysRevLett.75.890,PhysRevB.58.R1703, PhysRevB.98.075421, PhysRevA.92.063634, PhysRevLett.83.1275, PhysRevLett.96.210402, PhysRevLett.97.100402, PhysRevB.79.064409,PhysRevB.75.233104, PhysRevA.78.023631, PhysRevA.79.043633, Tang_2015,  PhysRevB.90.134201, PhysRevA.78.045602,PhysRevA.86.043631}. The exotic properties of 1D (Abelian) anyon models include  dynamical fermionization \cite{PhysRevA.78.045602,PhysRevA.86.043631,PhysRevA.96.023611, science367-1461}, asymmetric momentum distributions in ground state \cite{PhysRevB.79.064409, PhysRevB.75.233104, PhysRevA.78.023631, PhysRevA.79.043633, Tang_2015,Pu_2007, Calabrese_2009, PhysRevLett.118.120401,nc2-361}, anyonic symmetry protected topological phases \cite{PhysRevLett.118.120401}, entanglement properties \cite{PhysRevA.80.052332}, and statistics-induced Mott insulator to superfluid quantum phase transitions \cite{nc2-361, PhysRevLett.115.190402, PhysRevA.94.013611, PhysRevA.95.053614}. Several experimental schemes have been proposed for realizing anyonic statistics in ultracold atoms \cite{PhysRevLett.118.120401, nc2-361,PhysRevLett.115.053002, PhysRevLett.117.205303, PhysRevLett.121.030402} and photonic systems \cite{PhysRevA.96.043864} by engineering occupation-number dependent hopping using Raman-assisted tunneling or periodic modulation.

A recent surge of interest in the nonequilibrium dynamics of these 1D systems has been boosted by the powerful platform of cold atom systems \cite{AP56-243,RevModPhys.80.885} for simulating and probing nonequilibrium properties of quantum many-body systems \cite{nphys11-124,Gogolin_2016,PhysRevLett.110.205301,science353-794,PhysRevLett.119.080501}.
A paradigmatic model in this realm is the lattice hard-core anyons (HCAs) \cite{PhysRevLett.97.100402,PhysRevA.79.043633,PhysRevLett.113.050601}, which continuously interpolate between the noninteracting spinless fermions and hard-core bosons.
Yet, most of the nonequilibrium studies to date have focused on the quench dynamics of equal-time correlations, such as the density profile and the momentum distribution \cite{PhysRevA.78.045602,PhysRevA.86.043631,PhysRevA.95.063621, PhysRevA.96.023611, PhysRevLett.113.050601,science367-1461}, which can be obtained directly from the many-body wavefunction of the HCAs \cite{PhysRevLett.97.100402}. Only few studies have been devoted to the understanding of unequal-time (or, dynamical) correlations such as the Green's function and the out-of-time-ordered correlator (OTOC) \cite{PhysRevLett.126.065301,PhysRevLett.121.250404}.

The knowledge of such dynamical quantities have pivotal importance in characterizing the dynamical properties of the quantum system. Specifically, the Green's function and its spectral function allow to compute the signal of angle-resolved photoemission spectroscopy or momentum-resolved stimulated Raman spectroscopy, which have been performed in cold atom platforms \cite{Damascelli_2004,nature454-744,PhysRevLett.120.060404,PhysRevB.97.125117}. The OTOC has emerged as a diagnostic tool for chaos and information scrambling in quantum many-body systems \cite{PhysRevLett.115.131603, JHEP2016-1,JHEP2016-106, JHEP2016-129, JHEP2017-65, PhysRevLett.121.250404,PhysRevLett.126.030601,PhysRevA.103.062214}. It has also been applied to study a variety of many-body phenomena, ranging from quantum phase transitions \cite{PhysRevLett.123.140602} to many-body localization \cite{AdP529-1600318,AdP529-1600332,FAN2017707,PhysRevB.95.054201,PhysRevB.95.060201}.
One method of computing the Green's function of HCAs is to express it as a Fredholm determinant \cite{Zvonarev_2009,Pu_2008b,2110.06860}. However, this method is restricted to uniform systems and is difficult to be extended to arbitrary confining potential. In a recent Letter \cite{PhysRevLett.126.065301}, a general method has been developed to calculate the exact spectral function of 1D hard-core bosons for any type of confining potential, which makes use of the many-body wavefunction. However, it's challenging to extend it to finite temperatures or HCAs with arbitrary statistical angle.

In this work, we present a new approach to calculate the dynamical correlations of HCAs in one-dimensional lattices, without using the concrete form of the many-body wavefunction. Specifically, we provide an efficient method to compute dynamical correlations by using the basic properties of Gaussian operators and apply it to study the Green's function, the spectral function and the OTOC of 1D HCAs.
We find three main singularity lines in the spectral functions and obtain their dispersion relations by fitting the numerical results. The three lines cross at a common point $(q_m,\omega_m)$ where the spectral function reaches its largest value, and correspondingly the momentum distribution $n(q)$ of anyons exhibits a peak at $q_m$ while the local spectral function $A_{jj}(\omega)$ shows a power-law singularity at $\omega_m$. We also prove that the anyonic statistics can induce spatial asymmetry in the Green's function and its spectral function. Moreover, we diagnose information spreading by studying the OTOC, which shows asymmetric light-cone dynamics at low temperatures. However, as the temperature increases, the left and right butterfly velocities come close to each other and reach to the same value at infinite temperature.
Our results allow direct comparison with state-of-the-art experiments and provide a route to study the anyon non-equilibrium dynamics, especially to investigate the competing role of statistics, strong correlation, and external confining potential.


This paper is organized as follows. In Sec.\ref{sec:model} we give the model Hamiltonian of the hard-core anyons, and map it to a spiness fermion model by a generalized Jordan-Wigner transformation. In Sec.\ref{sec:dyn} we obtain explicit expressions for the Green's function, the spectral function and the OTOCs, and present results of numerical computations. In Sec.\ref{sec:symmetry} we analyze the symmetries of the dynamical correlation functions observed in the numerical results. We conclude in Sec.\ref{sec:conclusion} with a summary of our main results and some discussions. Some technical details are included in several Appendices.

\section{Model Hamiltonian} \label{sec:model}
We focus on the model of HCAs, which satisfy the generalized commutation relations \cite{PhysRevA.79.043633}
\begin{eqnarray}
&& \hat{a}_j\hat{a}_k^\dag + e^{-i\theta\, \text{sgn}(j-k)} \hat{a}_k^\dag \hat{a}_j =\delta_{jk}, \notag\\
&& \hat{a}_j\hat{a}_k + e^{i\theta\, \text{sgn}(j-k)} \hat{a}_k  \hat{a}_j =0,
\end{eqnarray}
where $\theta$ is the statistical parameter, $0\le \theta\le \pi$, and the sign function $\text{sgn}(x)=-1,0$ or $1$ depending on whether $x$ is negative, zero, or positive, respectively. When $j=k$, the commutation relations yield the hard-core constraints $\hat{a}_j^2=\hat{a}_j^{\dag 2}=0$ and $\{\hat{a}_j, \hat{a}_j^\dag\}=1$. Particularly, $\theta=0$ and $\theta=\pi$ correspond to spinless fermions and hard-core bosons, respectively, whereas for $0<\theta<\pi$ these commutations interpolate continuously between the two limiting cases.

We consider the dynamics of anyons confined in an optical lattice of $L$ sites, described by a tight-binding Hamiltonian
\begin{equation}\label{eq:Hamilton:HCA}
  \hat{H}= -J\sum_{j=1}^{L-1} (\hat{a}_{j}^\dag \hat{a}_{j+1}+\text{H.c.}) +\sum_{j=1}^L (V_j-\mu) \hat{n}_j,
\end{equation}
with a harmonic trap potential $V_j=\frac{1}{2}V_0^2[j-(L+1)/2]^2$, where $V_0$ denotes the strength of the trap. However, we stress that this special form of potential is not necessary since our formalism developed in paper is valid for any type of confining potential, even random $V_j$'s. The chemical potential $\mu$ is included to control the filling factor. Hereafter, we work in units where hopping parameter $J=\hbar=1$. By a generalized Jordan-Wigner transformation \cite{PhysRevLett.97.100402,PhysRevA.79.043633,PhysRevA.86.043631}
\begin{equation}\label{eq:JW}
  \hat{a}_j= e^{-i\theta\sum_{l< j}\hat{c}_l^\dag\hat{c}_l} \hat{c}_j, \quad
  \hat{a}_j^\dag=  \hat{c}_j^\dag e^{i\theta\sum_{l< j}\hat{c}_l^\dag\hat{c}_l},
\end{equation}
where $\hat{c}_j^\dag (\hat{c}_j)$ are creation (annihilation) operators for spinless fermions, the hard-core anyon Hamiltonian can be mapped to a spinless fermion Hamiltonian
\begin{equation}\label{eq:Hamilton:SF}
  \hat{H}_F= -\sum_{j=1}^{L-1} (\hat{c}_{j}^\dag \hat{c}_{j+1}+\text{H.c.}) +\sum_{j=1}^L (V_j-\mu) \hat{n}_j,
\end{equation}
where $\hat{n}_j=\hat{c}_j^\dag\hat{c}_j=\hat{a}_j^\dag\hat{a}_j$. This Hamiltonian is a bilinear form of the fermion creation and annihilation operators, which can be written as $\hat{H}_F=\sum_{lm} \hat{c}_l^\dag \mathds{H}_{lm}\hat{c}_m$, where $\mathds{H}$ is an $L\times L$ matrix, with matrix elements $\mathds{H}_{lm}=-\delta_{l,m\pm1} +(V_l-\mu)\delta_{lm}$.

\section{Dynamical Correlations}\label{sec:dyn}
\subsection{Single-Particle Green's Function}
Now consider the single-particle Green's functions in a thermal state described by the density matrix
$\hat{\rho}=e^{-\beta\hat{H}}/\text{Tr}[e^{-\beta\hat{H}}]$, where $\beta$ is the inverse temperature, $\beta=1/(k_BT)$.
We define the lesser and greater Green's functions of the hard-core anyons as
\begin{equation}\label{eq:GF:definition}
G^{<}_{jk}(t)\equiv -i\langle \hat{a}_k^\dag \hat{a}_j(t)\rangle, \quad  G^{>}_{jk}(t)\equiv -i\langle  \hat{a}_j(t) \hat{a}_k^\dag \rangle,
\end{equation}
where $\langle\hat{O}\rangle\equiv \text{Tr}[\hat{\rho}\hat{O}]$. For $\theta=0$ and $\theta=\pi$, these Green's functions reduce to that of spinless fermions and hard-core bosons, respectively. Other types of nonequilibrium Green's functions can be expressed in terms of $G^{\gtrless}(t)$ and hence it's sufficient to analyze the properties of these two Green's functions.

Employing the basic properties of the Gaussian operators together with the generalized Jordan-Wigner transformation, we obtain the following explicit expressions for the Green's functions, which constitute one of our main results [see Appendix.\ref{sec:GF:append} for some details]:
\begin{subequations}
\begin{equation}\label{eq:GF:sol:lesser}
  iG^{<}_{jk}(t)
=\det\left[ \tilde{\mathds{B}}^{jk}(t)\right] \left\{ e^{-it\mathds{H}} (\mathds{1}-\mathds{B}_0)\left[\tilde{\mathds{B}}^{jk}(t)\right]^{-1}  \right\}_{jk},
\end{equation}
\begin{equation}\label{eq:GF:sol:greater}
iG^{>}_{jk}(t)
= \det\left[  \mathds{B}^{jk}(t) \right] \left\{  e^{-it\mathds{H}} \mathds{B}_0
\left[\mathds{B}^{jk}(t)\right]^{-1}  \mathds{P}^{j}_{-}(t) \right\}_{jk},
\end{equation}
\end{subequations}
where $\mathds{1}$ denotes the $L$-dimensional identity matrix, $\mathds{B}_0=[\mathds{1}+e^{-\beta\mathds{H}}]^{-1}$, $\mathds{P}^{j}_{\pm}(t)=e^{it\mathds{H}} e^{\pm\mathds{J}^{(j)}} e^{-it\mathds{H}}$, and
\begin{eqnarray*}
&& \mathds{B}^{jk}(t)=\mathds{B}_0+ \mathds{P}^{j}_{-}(t) \mathds{P}^{k}_{+}(0)(\mathds{1}-\mathds{B}_0), \\
&& \tilde{\mathds{B}}^{jk}(t)=\mathds{B}_0+ \mathds{P}^{k}_{+}(0) \mathds{P}^{j}_{-}(t) (\mathds{1}-\mathds{B}_0).
\end{eqnarray*}
The matrix $\mathds{J}^{(j)}$ is a diagonal $L\times L$ matrix with $\mathds{J}^{(j)}_{ll}$ equals to $i\theta$ for $l<j$ and 0 for $l\ge j$. The matrix $\mathds{B}_0$ is the static correlation function of the mapped fermions in the thermal state, with matrix elements $(\mathds{B}_0)_{jk}=\langle \hat{c}_j\hat{c}_k^\dag\rangle$.

Obviously, when $\theta=0$, $\mathds{P}^{j}_{\pm}(t)=\mathds{B}^{jk}(t)=\tilde{\mathds{B}}^{jk}(t)=\mathds{1}$, and hence the above expressions for the Green's function reduce to the results of the free spinless fermions.
The expression for the lesser Green's function in Eq.(\ref{eq:GF:sol:lesser}) also contains as a limiting case the result for the one-body density matrix at equal times, $\rho_{kj}=iG^{<}_{jk}(t=0)$.
Most importantly, although the above expressions are given on a finite lattice, they can also be used to obtain the Green's functions of HCAs in continuous space with discrete single-particle spectrum [see Appendix.\ref{sec:lattice:continuous}]. Remarkably, for the special case of $\theta=\pi$ (hard-core bosons, or Tonks-Girardeau gas) and zero temperature, our result is essentially equivalent to the expressions given in a recent Letter \cite{PhysRevLett.126.065301}. However, we stress that Eqs.(\ref{eq:GF:sol:lesser}) and (\ref{eq:GF:sol:greater}) are valid for any temperature $T$, any statistical angle $\theta$, and any trap potential $V_j$.

\begin{figure}
  \centering
  \includegraphics[width=0.48\textwidth]{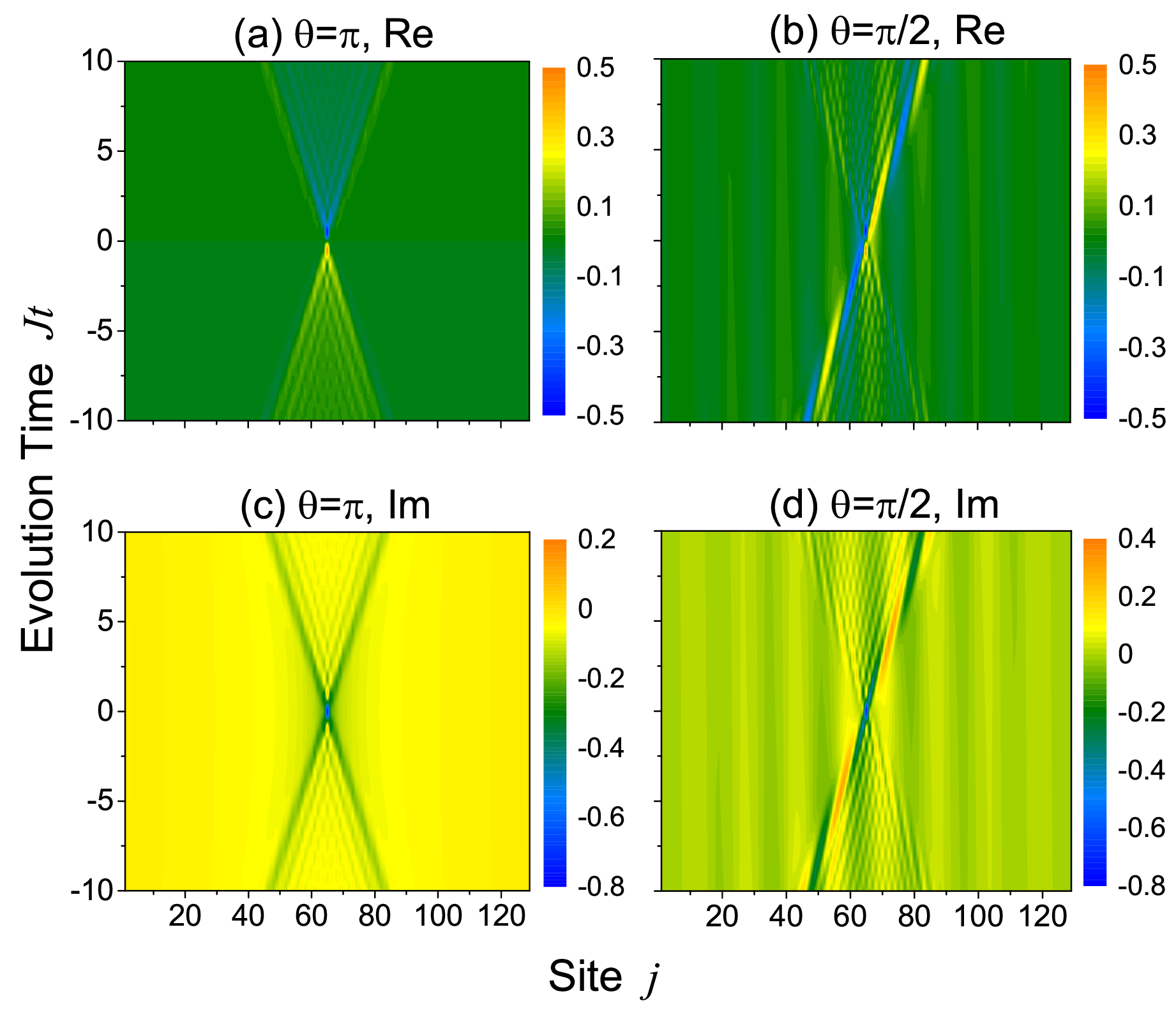}\\
  \caption{The real (top panel) and imaginary (bottom panel) part of the greater Green's function $G^{>}_{jk}(t)$ in real space-time for $\theta=\pi$ (left panel) and $\theta=\pi/2$ (right panel). The system parameters: the temperature $T=0$, the potential strength $V_0=0$, the chemical potential $\mu=-1.4$, the chain length $L=129$, and the site $k$ is fixed at the middle, $k=65$.} \label{fig:GF}
\end{figure}
The expressions for the lesser and greater Green's functions [Eqs.(\ref{eq:GF:sol:lesser}) and (\ref{eq:GF:sol:greater})] are especially suitable for numerical computations since only linear algebra is needed. Fig.\ref{fig:GF} shows the numerical results of the greater Green's function $G^{>}_{jk}(t)$ in real space-time for $\theta=\pi, \pi/2$ and fixed $k=65$ in a lattice with $L=129$.
We see that the propagation of the single-particle excitation exhibits a clear symmetric light-cone for $\theta=\pi$. However, the propagation is asymmetric for $\theta=\pi/2$, as shown in Figs.\ref{fig:GF}(b) and \ref{fig:GF}(d). This spatial asymmetry is a general feature for $\theta\neq0,\pi$, caused by the novel statistics of anyons, and can also exist in other dynamical correlation functions. A symmetry analysis about the dynamical correlations would be given in Sec.\ref{sec:symmetry}.
We comment here that both the real and imaginary parts of the Green's function are important in real space. This is because (i) both of them are necessary in analyzing the spatial symmetry [see Sec.\ref{sec:symmetry}], and (ii) the Green's function in real space reflects the propagation amplitude of one-particle excitations,  and hence both the real and imaginary parts have physical relevance, in analogy to the physics of a wavefunction.

\subsection{Spectral Function}
From the above Green's functions in real space-time one can define two spectral functions
\begin{equation}\label{eq:spectral:real-space}
  A_{jk}^{\pm}(\omega)=\frac{i}{2\pi} \int_{-\infty}^\infty {G}_{jk}^{\gtrless}(t)\, e^{i\omega t} dt.
\end{equation}
They are related by $A^{-}_{jk}(\omega)=e^{-\beta\omega} A^{+}_{jk}(\omega)$ at finite temperature $T=(k_B\beta)^{-1}$.
Transforming to momentum space, we have
\begin{equation}
  A^{\pm}(q,\omega)\equiv \frac{1}{L}\sum_{jk} A_{jk}^{\pm}(\omega) e^{-iq(j-k)}.
\end{equation}
In Appendix.\ref{sec:GF:analytic} we prove that both $A^{\pm}(q,\omega)$ and the local spectral function $A_{jj}^{\pm}(\omega)$ are nonnegative real numbers, and hence have probability-density interpretation. For example,  $A^{\pm}(q,\omega)$ correspond to the probability density for a particle (hole) to be excited (filled) at a given momentum $q$ and energy $\omega$.

\begin{figure}
  \centering
  \includegraphics[width=0.48\textwidth]{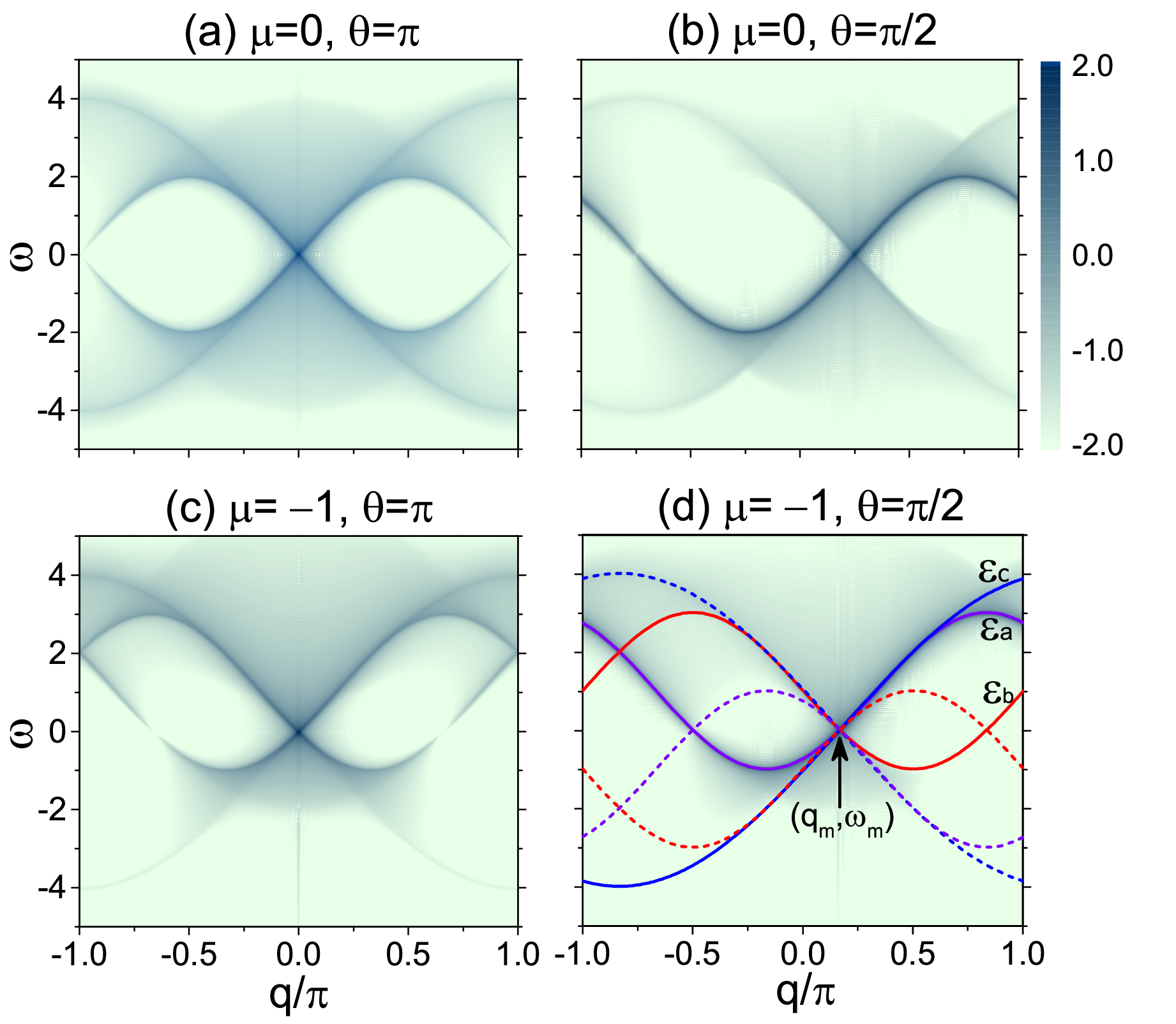}\\
  \caption{(Color online) Logarithm of the spectral function, $\log_{10} A(q,\omega)$, of the hard-core anyons on a lattice in the $(q,\omega)$ plane for different statistical angle $\theta$ and chemical potential $\mu$. Here $V_0=0, T=0$, and $L=128$. In (d), violet, red and blue solid lines mark the excitation singularities $\varepsilon_a(q), \varepsilon_b(q)$ and $\varepsilon_c(q)$, respectively, while the corresponding dashed lines mark $-\varepsilon_a(q), -\varepsilon_b(q)$ and $-\varepsilon_c(q)$.} \label{fig:GF-Spectrum}
\end{figure}

Figure \ref{fig:GF-Spectrum} shows the total spectral function $A(q,\omega)=A^{+}(q,\omega)+A^{-}(q,\omega)$ for various chemical potential $\mu$ and statistical angle $\theta$ at $T=0$ in a finite lattice with $V_0=0$.
The $\omega\ge0 (\le0)$ part of $A(q,\omega)$ comes from the greater (lesser) Green's functions. Two special values of $\mu=0,-1$ are chosen in the numerical plot, however, the features discussed below are quite general.
The spectral function strongly depends on the statistical angle $\theta$ and the chemical potential $\mu$, or equivalently, the filling factor $\nu=q_F/\pi$, with $q_F$ being the Fermi wavevector of the mapped spinless fermion. There are three pairs of main singularity lines [see Fig.\ref{fig:GF-Spectrum}(d)], denoted as $\pm\varepsilon_a(q), \pm\varepsilon_b(q)$ and $\pm\varepsilon_c(q)$. By fitting the numerical results for various $\mu$ and $\theta$, we find that
$\varepsilon_a(q)=-2\cos(q+\nu\theta)-\mu$,
$\varepsilon_b(q)=-2\cos(q+\nu\theta-2\nu\pi)-\mu$, and
$\varepsilon_c(q)=4\sin[(q+\nu\theta-\nu\pi)/2]$.
When $\theta=0$, the spectral weight lies completely on the dispersive curve $\varepsilon_a(q)=-2\cos(q)-\mu$, as it should be for noninteracting fermions. However, as $\theta$ increases, the string operator $e^{i\theta\sum_{l< j}\hat{c}_l^\dag\hat{c}_l}$ in the anyon creation operator $\hat{a}_j^\dag$ may induce two effects: (i) the excitation singularity lines are momentum-shifted, $q\rightarrow q+\nu\theta$; (ii) the spectral weight is transferred from $\varepsilon_a(q)$ to other singularity lines, due to the particle-hole excitations induced by the string operator.

Physically, we can understand these singularity lines as a double spectrum, similar to that of an interacting Bose gas \cite{PhysRev.130.1616}. 
In detail, the first dispersion line $\varepsilon_a(q)$ could be identified as the anyonic  analogue of Lieb-I modes of the Lieb-Liniger gas, corresponding to a particle with momentum $q_F$ promoted to a generic state with momentum $q+\nu\theta$. The second line $\varepsilon_b(q)$ corresponds to a particle with momentum $q_F$ promoted to a  state with momentum $q+\nu\theta-2 q_F$.
The third line $\varepsilon_c(q)$ corresponds to a symmetric excitation of a particle from an occupied state at momentum $\pi/2-(q+\nu\theta-q_F)/2$ to a free one with momentum $\pi/2+(q+\nu\theta-q_F)/2$, in analogy to the hard-core boson in a lattice \cite{PhysRevLett.126.065301}, but without any analogue in the homogeneous case.

The momentum shift $q\rightarrow q+\nu\theta$ observed above can be understood qualitatively in a mean-field manner. In the language of the mapped spinless fermion, the anyon excitation is $\hat{a}_j^\dag=\hat{c}_j^\dag e^{i\theta\sum_{l< j}\hat{c}_l^\dag\hat{c}_l}$. In mean-field approximation, $\hat{a}_j^\dag\approx\hat{c}_j^\dag e^{i\theta \nu(j-1)}$, and hence $\hat{a}_q^\dag\sim \hat{c}_{q+\nu\theta}^\dag$, resulting in a momentum shift $q\rightarrow q+\nu\theta$ in the dispersion relations.

When $\theta=\pi$, the two lines $\varepsilon_a(q)$ and $\varepsilon_b(q)$ have the same weight since $A(q,\varepsilon_a)=A(-q,\varepsilon_b)$ [see Figs.\ref{fig:GF-Spectrum}(a) and \ref{fig:GF-Spectrum}(c)], which is a result of the symmetry property $A(q,\omega)=A(-q,\omega)$ and $\varepsilon_a(q)=\varepsilon_b(-q)$ for $\theta=\pi$.
However, we should note that the Green's function and hence its spectral function have no spatial inversion symmetry for $\theta\neq0,\pi$ [see Figs.\ref{fig:GF-Spectrum}(b) and \ref{fig:GF-Spectrum}(d)], although the Hamiltonian is invariant under the reflection about the middle of the chain. We would show in Sec.\ref{sec:symmetry} that $A^{\pm}(q,\omega;\theta)=A^{\pm}(-q,\omega;-\theta)$, where the spectral function is labeled with the sign of the statistical parameter for convenience. Then the spectral function has spatial inversion symmetry only for $\theta=0$ (spinless fermions) and $\theta=\pi$ (hard-core bosons). This asymmetry may provide us a qualitative approach for detecting anyonic statistics by using dynamical correlations in ultracold atom systems.
Since anyonic statistics have been proposed to be realizable in ultracold atoms \cite{PhysRevLett.118.120401, nc2-361,PhysRevLett.115.053002, PhysRevLett.117.205303, PhysRevLett.121.030402} and spectral functions could be measured in cold atom platforms \cite{Damascelli_2004,nature454-744,PhysRevLett.120.060404,PhysRevB.97.125117}, we expect that the spectral function of anyons is accessible to current state-of-the-art experiments with ultracold atoms.

\begin{figure}
  \centering
  \includegraphics[width=0.48\textwidth]{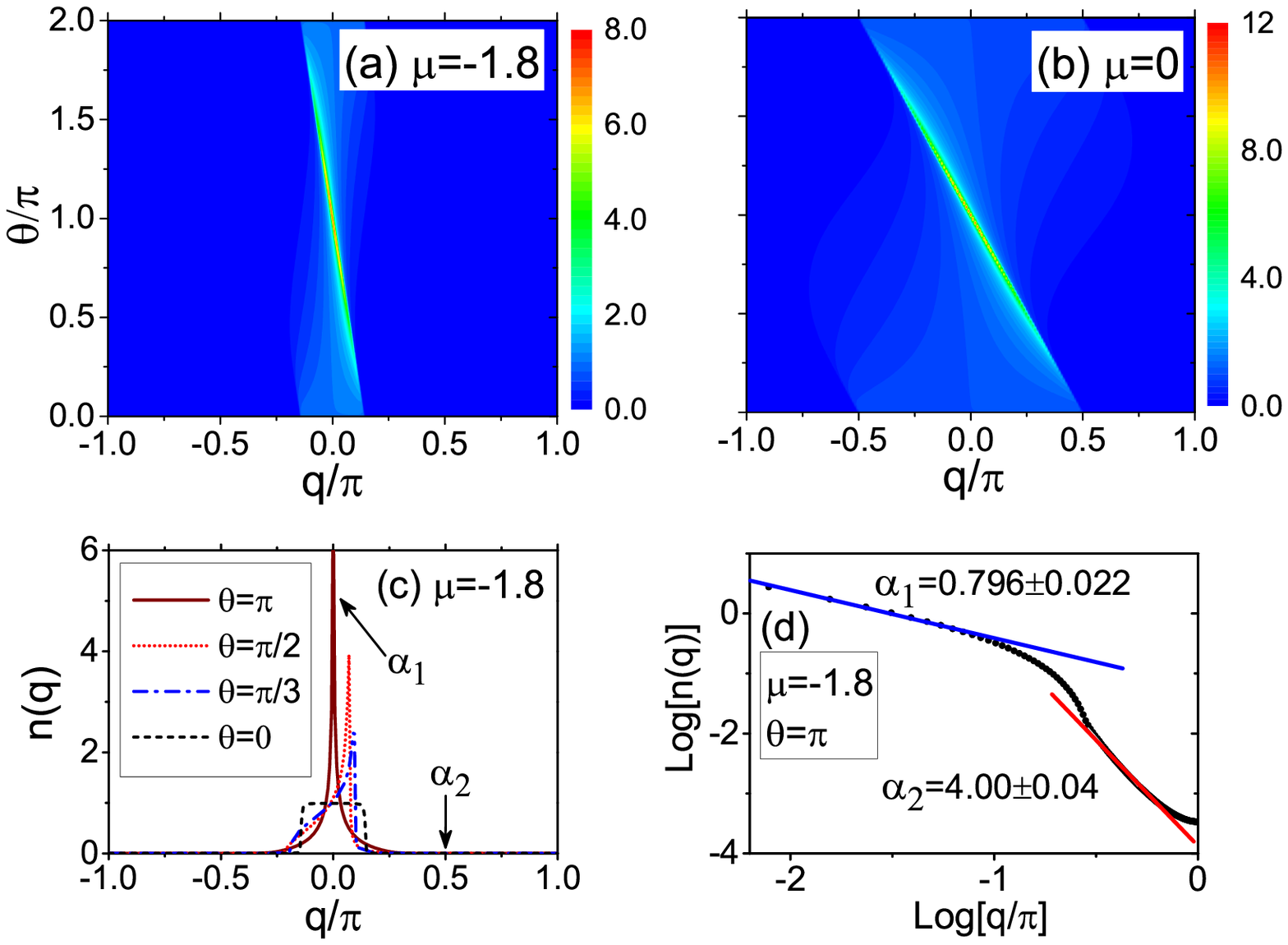}\\
  \caption{The ground-state momentum distribution $n(q)$ of anyons in a lattice with $L=256$ sites and $V_0=0$. (a)-(b) The distribution as a function of $q$ and $\theta$ for $\mu=-1.8 (\nu\approx 0.14)$ and $\mu=0 (\nu=0.5)$.
  (c) Cut of (a) for several representative statistical angles $\theta=0, \pi/3,\pi/2$ and $\pi$. For $\theta=\pi$, the distribution $n(q)$ shows power law behavior: $n(q)\sim q^{-\alpha_1}$ near the central peak, while $n(q)\sim q^{-\alpha_2}$ in the high-momentum regime, as shown in (d) with fitted values of the exponents $\alpha_{1,2}$. }\label{fig:dist}
\end{figure}

A remarkable feature of the spectral function $A(q,\omega)$ is that the three dispersion lines cross each other at a ``\emph{hot  point}'' $(q_m,\omega_m)$, as shown in Fig.\ref{fig:GF-Spectrum}(d), near which the spectral weight is largest in the whole $(q,\omega)$ plane. From the explicit expressions of the three lines we obtain
\begin{equation}\label{eq:hotPoint}
  q_m=\nu(\pi-\theta), \quad \omega_m=0.
\end{equation}
This hot point and its linear dependence on $\nu$ and $\theta$ may provide us an exact quantitative experimental signature to probe anyonic statistics through nonequilibrium dynamics.

Since our calculation is exact within the numerical  accuracy at all energy scales, we can check the sum rules satisfied by  the spectral functions $A^{\pm}(q,\omega)$. Especially, integration over all frequencies of $A^{-}(q,\omega)$ gives the momentum distribution $n(q)$. Therefore the hot point should correspond to a peak in the function $n(q)$ for $\theta\neq0$, which is indeed the case as shown in Fig.\ref{fig:dist}.
Figures \ref{fig:dist}(a) and (b) show that the momentum $q_m$ indeed is a linear function of $\theta$ and is exactly given by Eq.(\ref{eq:hotPoint}). 
Fig.\ref{fig:dist}(c) plots $n(q)$ for several representative statistical angles. We see that $n(q)$ is the well-known Fermi-Dirac distribution for $\theta=0$, while for $0<\theta\le\pi$ it has a peak at $q_m$. For $\theta=\pi$, $n(q)$ displays power-law behaviors: $n(q)\sim q^{-\alpha_1}$ in the small momentum regime ($q\rightarrow0$) \cite{scipost8-053}, and  $n(q)\sim q^{-\alpha_2}$ in the ``high momentum regime'' (i.e., the regime where $q$ is far from both 0 and $\pi$), as shown in Fig.\ref{fig:dist}(d). The fitted exponent $\alpha_2\approx 4$, consistent with the universal $q^{-4}$ tail in the momentum distribution of the Lieb-Liniger gas \cite{PhysRevLett.91.090401}. We remark that to show the high-momentum $q^{-4}$ tail clearly, the filling factor should be small enough. In Figs.\ref{fig:dist}(c)-(d) we choose $\mu=-1.8$ to give a relatively low filling factor $\nu\approx 0.14$.

\begin{figure}
  \centering
  \includegraphics[width=0.48\textwidth]{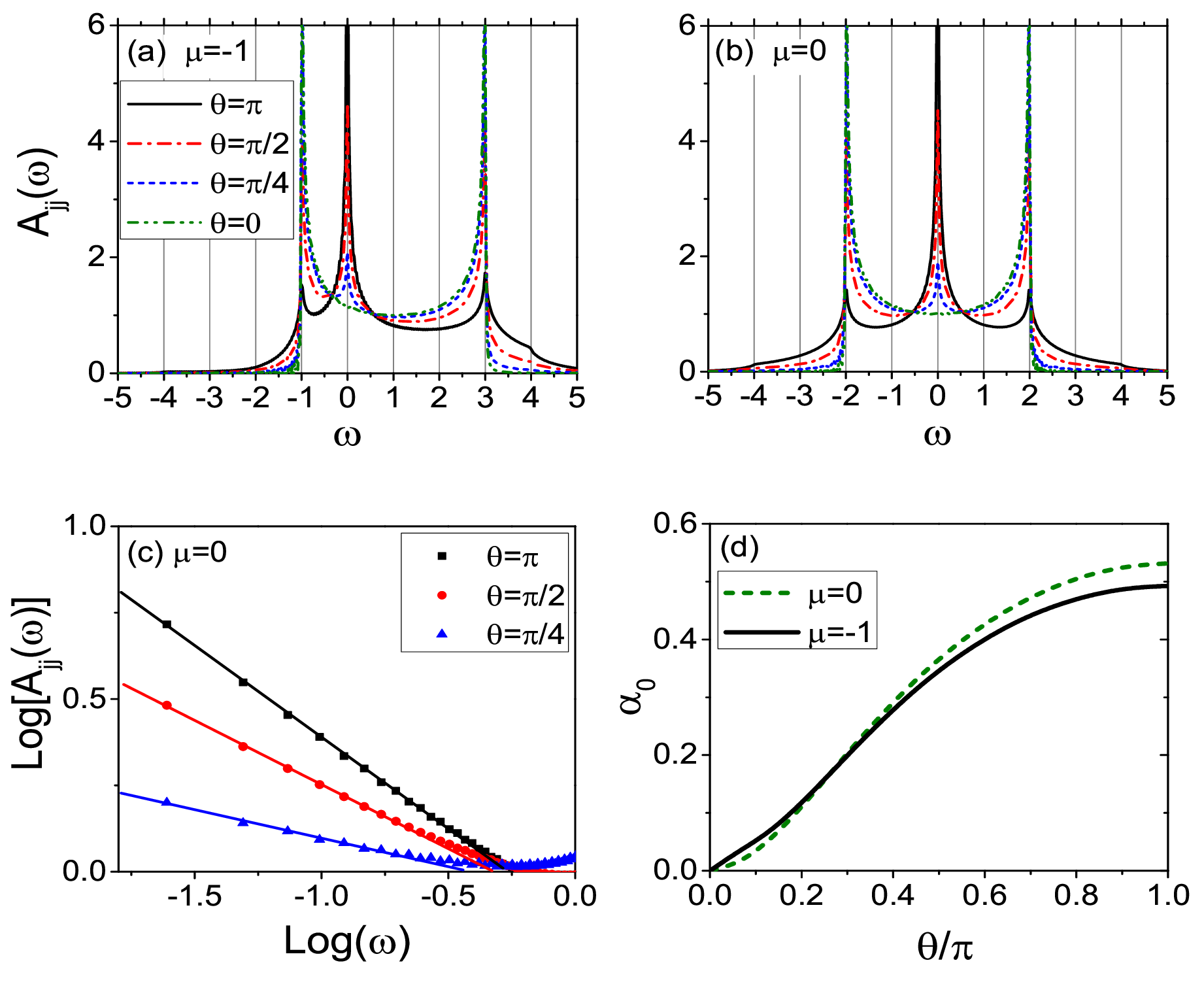}\\
  \caption{(a)-(b) The local spectral function $A_{jj}(\omega)$ for $\mu=-1,0$ and $j=256$ in a lattice with $L=512$. (c) The power law behavior near $\omega_0=0$ for $\mu=0$: $A_{jj}(\omega)\propto |\omega-\omega_{0}|^{-\alpha_0}$. (d) The fitted exponent $\alpha_0$ as a function of $\theta$ for two different chemical potentials, $\mu=0$ and $\mu=-1$.  Other parameters: the temperature $T=0$ and the potential strength $V_0=0$.  }\label{fig:LDOS}
\end{figure}
Furthermore, the local spectral functions $A_{jj}^{\pm}(\omega)$ are also important observables in some experiments such as the scanning tunneling microscopy\cite{PhysRevA.76.063602,PhysRevX.8.011037}. In our formalism they are even easier to compute than $A^{\pm}(q,\omega)$ since the basic equations (\ref{eq:GF:sol:lesser}) and (\ref{eq:GF:sol:greater}) are written in real space and time. Figures \ref{fig:LDOS}(a) and \ref{fig:LDOS}(b) show some examples of $A_{jj}(\omega)=A^{+}_{jj}(\omega)+ A^{-}_{jj}(\omega)$ for various statistical parameters and chemical potentials in a lattice with $L=512$. The spectrum shows singularities for $\theta\neq0$ at five critical frequencies: $\omega_0=0,\omega_{ab}^{\pm}=\pm2-\mu$ and $\omega_c^{\pm}=\pm4$. This could be understood from the structure of $A(q,\omega)$. The singularity at $\omega_0=0$ comes from the hot point $(q_m,\omega_m)$, which is strongest for $\theta=\pi$ and vanishes for $\theta=0$. The critical frequencies $\omega_{ab}^{\pm}$ and $\omega_{c}^{\pm}$ correspond to the top and bottom of the dispersion curves $\varepsilon_{a,b}(q)$ and $\varepsilon_{c}(q)$, respectively. At $\omega_0=0$ and $\omega_{ab}^{\pm}$ the local spectral function diverges but at $\omega_c^{\pm}$ there is no divergence.
According to the nonlinear Luttinger liquid theory \cite{science323-228,RevModPhys.84.1253}, the divergence near $\omega_0$ and $\omega_{ab}^{\pm}$ should show power law behavior, $A_{jj}(\omega)\propto |\omega-\omega_{j}|^{-\alpha_j}$, $j=0$ or $ab$. We analyze this power-law behavior in detail for the singularity at $\omega_0$ in Figs.\ref{fig:LDOS}(c) and \ref{fig:LDOS}(d). We see that this exponent increases monotonically with the statistical parameter $\theta$. When $\theta=0$, $\alpha_0=0$ since there is no divergence at all. On the other hand, when $\theta=\pi$, $\alpha_0$ should be equal to the exponent of the singularity near the hot point, which is $1/2$ according to the mobile impurity theory \cite{PhysRevLett.100.206805,science323-228,PhysRevLett.102.126405,RevModPhys.84.1253,scipost3-015}. The numerical results  shown in Fig.\ref{fig:LDOS}(d) are close, but not exactly coinciding with the predicted value $\alpha_0|_{\theta=\pi}=1/2$.
This difference is expected because we consider a finite lattice rather than a homogeneous system and go beyond the approximations used in the phenomenological theory.

\begin{figure}
  \centering
  \includegraphics[width=0.48\textwidth]{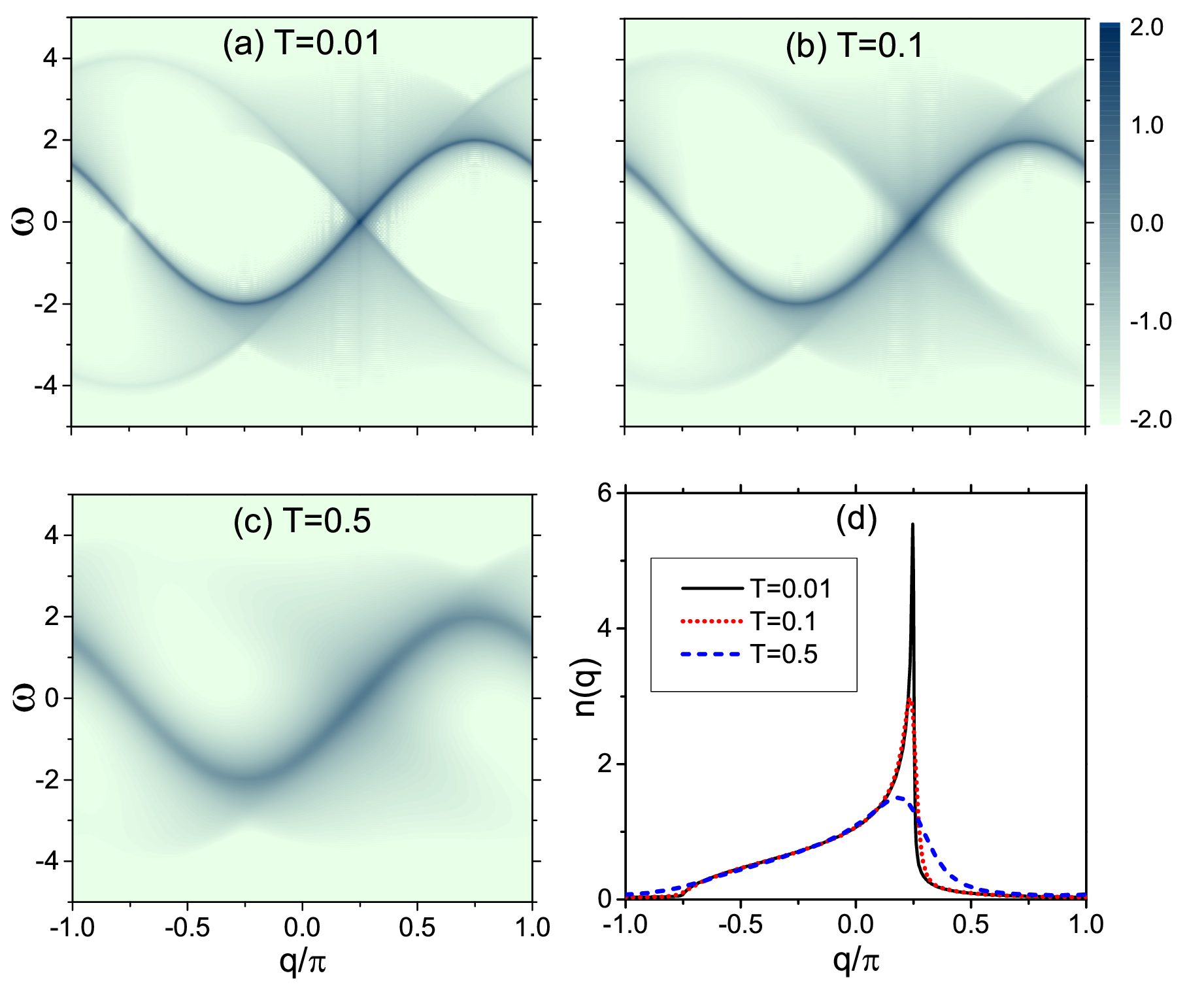}\\
  \caption{(a)-(c) (Color online) Logarithm of the spectral function, $\log_{10} A(q,\omega)$, of the hard-core anyons on a lattice in the $(q,\omega)$ plane at finite temperatures $T$. Here $\theta=\pi/2,\mu=0, V_0=0$ and $L=128$. (d) The momentum distribution $n(q)$ of anyons at finite temperatures for $\theta=\pi/2,\mu=0, V_0=0$ and $L=256$. }\label{fig:spec:T}
\end{figure}
So far we have focused on the zero-temperature properties. However, the Eqs.(\ref{eq:GF:sol:lesser}) and (\ref{eq:GF:sol:greater}) are valid for any temperature and let's now discuss the impact of temperature briefly. Figures \ref{fig:spec:T}(a)-\ref{fig:spec:T}(c) show the spectral function $A(q,\omega)$ at finite temperatures for $\theta=\pi/2$ and $\mu=0$. Comparing with the zero-temperature result shown in Fig.\ref{fig:GF-Spectrum}(b) we can observe the main effect of finite temperature: the singularities at the dispersion lines $\pm\varepsilon_{a,b,c}$ are suppressed and broadened as the temperature increases. This is because the thermal fluctuations would destroy the coherence of anyonic excitations. The suppression and broadening effect could also be observed in the momentum distribution function $n(q)$ as shown in Fig.\ref{fig:spec:T}(d). We can check that this is a general feature for arbitrary statistical parameter $0<\theta\le\pi$ and chemical potential $\mu$. The noninteracting fermion ($\theta=0$) case is special: the spectral function does not broaden as the temperature increases, but the momentum distribution function broadens.

\subsection{Out-of-time-ordered correlator}
One advantage of our method is that it can be used to compute not only the two-point Green's functions but also any $n$-point dynamical correlations. Here we study an important dynamical quantity, the so-called OTOC, which can characterize the information spreading in an interacting quantum many-body system and has received tremendous interest \cite{nphys11-124,PhysRevLett.121.250404,nature481-484,PhysRevB.96.020406,PhysRevB.96.054503, Bohrdt_2017,scipost9-024,PhysRevLett.127.070403}.
The information spreading usually occurs in a spatially symmetric way for conventional fermionic or bosonic systems with translation invariance. However, this is not generally the case for anyonic systems, where statistics can induce asymmetric spreading of quantum information \cite{PhysRevLett.121.250404,scipost9-024}.

We define the anyonic OTOC as
\begin{equation}\label{eq:otoc:definition}
  F_{jk}(t)=\langle \hat{a}_j^\dag(t) \hat{a}_k^\dag(0) \hat{a}_j(t) \hat{a}_k(0)\rangle.
\end{equation}
Another main result in this work is the explicit expression for this OTOC [see Appendix.\ref{sec:OTOC:append} for some details]:
\begin{widetext}
\begin{equation}\label{eq:otoc:sol}
F_{jk}(t)= \det[\mathds{C}^{jk}(t)]  \left\{ \left[ e^{-it\mathds{H}} e^{-\mathds{J}^{(k)}} \mathds{Q}^{jk}(t) \mathds{P}^{j}_{+}(t) \right]_{jk} \left[\mathds{Q}^{jk}(t) e^{it\mathds{H}} \right]_{kj} -\left[ e^{-it\mathds{H}} e^{-\mathds{J}^{(k)}} \mathds{Q}^{jk}(t) e^{it\mathds{H}}  \right]_{jj} \left[\mathds{Q}^{jk}(t) \mathds{P}^{j}_{+}(t) \right]_{kk}  \right\},
\end{equation}
\end{widetext}
where $\mathds{C}^{jk}(t) \equiv \mathds{B}_0+\mathds{M}^{jk}(t) (\mathds{1}-\mathds{B}_0)$, $\mathds{Q}^{jk}(t) \equiv (\mathds{1}-\mathds{B}_0) \left[ \mathds{C}^{jk}(t)\right]^{-1}$, and
$\mathds{M}^{jk}(t)\equiv \mathds{P}^{j}_{+}(t) e^{\mathds{J}^{(k)}} \mathds{P}^{j}_{-}(t) e^{-\mathds{J}^{(k)}}$  are all $L\times L$ matrices.

\begin{figure}
  \centering
  \includegraphics[width=0.48\textwidth]{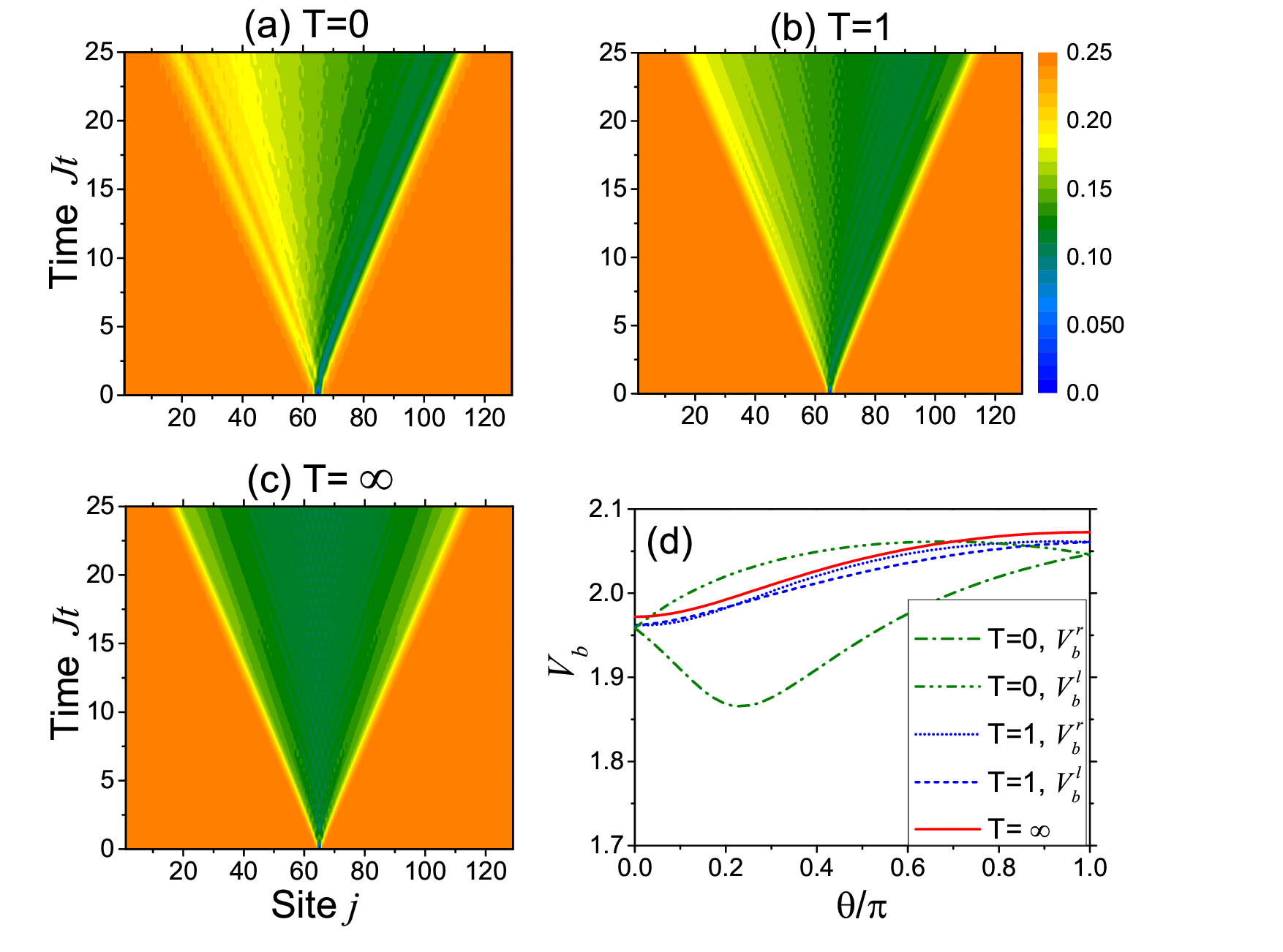}\\
  \caption{(a)-(c) OTOC growth $|F_{jk}(t)|$ for statistical angle $\theta=\pi/2$ and three different temperatures. Here $V_0=0,\mu=0,L=129$ and $k=65$. (d) Left ($V_b^l$) and right ($V_b^r$) butterfly velocities' dependence on the statistical angle $\theta$ for three different temperatures: $T=0, 1$ and infinity. }\label{fig:OTOC}
\end{figure}

Figures \ref{fig:OTOC}(a)-\ref{fig:OTOC}(c) show numerical results for $\theta=\pi/2$ and various temperatures. For free spinless fermions ($\theta=0$) and hard-core bosons ($\theta=\pi$), the OTOCs map out a symmetric light cone, which can be proved by symmetry analysis. However, for anyons ($\theta\neq0,\pi$) and low temperatures the information propagation is spatially asymmetric, as shown in Figs.\ref{fig:OTOC}(a)-\ref{fig:OTOC}(b). This asymmetry is suppressed as the temperature increases and vanishes at infinite temperature, as shown in  Fig.\ref{fig:OTOC}(c).
Physically, this is because that the asymmetry varies with the eigenstates of the Hamiltonian and hence at finite temperatures the thermal fluctuations would average the OTOC $F_{jk}(t)$ among different eigenstates and suppress the asymmetry. At infinite temperature the asymmetry would be averaged to zero [see Sec.\ref{sec:symmetry} for a proof based on symmetry analysis].
In addition, we comment here that the OTOCs can be observed using state-of-the-art technologies in the ground-state \cite{PhysRevLett.121.250404} or at finite temperatures\cite{PhysRevLett.128.140601}, and hence our discussions are experimentally relevant.

To further illustrate the OTOC's growth for right and left propagation directions, we plot the butterfly velocities in Fig.\ref{fig:OTOC}(d). We define a butterfly velocity $V_b$ by the boundary of the space-time region where  $|F_{jk}(t)|$ is suppressed by at least 1\% of its initial value \cite{PhysRevLett.121.250404}. As shown by the results, the left information propagation velocity is always larger than the right one for $T=0$ and $0<\theta<\pi$. However, as the temperature increases, the two velocities come close to each other and finally converge to the same value at infinite temperature.

\section{Symmetry Analysis}\label{sec:symmetry}
We have seen that the Green's function [Figs.\ref{fig:GF}(b) and \ref{fig:GF}(d)], the spectral function [Figs.\ref{fig:GF-Spectrum}(b) and \ref{fig:GF-Spectrum}(d)], the momentum distribution function [Figs.\ref{fig:dist}(a)-(c)], and the low temperature OTOC [Figs.\ref{fig:OTOC}(a)-(b)] do not have spatial inversion symmetry for $\theta\neq 0,\pi$, although the Hamiltonian is invariant under spatial reflection about the middle of the chain. This is in sharp contrast to the properties of conventional fermions or bosons.
To understand this problem we focus on the symmetry properties of the mapped free fermion model, i.e., $\hat{H}_F$ given by Eq.(\ref{eq:Hamilton:SF}). We would label physical quantities with the statistical parameter $\theta$ for convenience.

Consider the spatial inversion $\mathcal{I}$ under which the site $j$ is mapped to $j'=L+1-j$, and $\hat{c}_j \rightarrow \hat{c}_{j'}, \hat{c}_j^\dag \rightarrow \hat{c}_{j'}^\dag$.
Then it's straightforward to show that the Green's functions satisfy
\begin{eqnarray}\label{eq:inversion:GF}
   G^{\gtrless}_{jk}(t;\theta) &=& G^{\gtrless}_{j', k'}(t;-\theta), \\
  G^{\gtrless}(q,t;\theta) &=& G^{\gtrless}(-q,t;-\theta),
\end{eqnarray}
with the corresponding spectral functions $A^{\pm}(q,\omega;\theta)=A^{\pm}(-q,\omega;-\theta)$. Similarly the OTOC satisfies $F_{jk}(t;\theta)=F_{j'k'}(t;-\theta)$. Therefore, the Green's function, the spectral function and the OTOC are asymmetric except for the two special cases $\theta=0$ and $\theta=\pi$.

The Hamiltonian is also invariant under the time-reversal operator $\mathcal{T}$, which acts by complex-conjugating a state or operator written in the fermionic Fock basis. Using this time-reversal symmetry we can show that $G_{jk}^{\gtrless}(t; \theta) =G_{kj}^{\gtrless}(t; -\theta)$ and $F_{jk}(t;\theta) =F_{kj}(t;-\theta)$. By combining the two operators $\mathcal{I}$ and $\mathcal{T}$, we have
\begin{eqnarray}
G_{jk}^{\gtrless}(t; \theta) &=& G_{k'j'}^{\gtrless}(t; \theta), \\
F_{jk}(t;\theta) &=&F_{k'j'}(t;\theta). \label{eq:OTOC:inversion:2}
\end{eqnarray}
From the first equation we can conclude that the local Green's function $G_{jj}^{\gtrless}(t)$ [and the local spectral function $A_{jj}^{\pm}(\omega)$] is symmetric in real space, i.e., $G_{jj}^{\gtrless}(t)=G_{j'j'}^{\gtrless}(t)$. This is in contrast to the property of anyons with finite interaction \cite{PhysRevLett.121.250404}.

At infinite temperature, additional symmetries of the dynamical correlations may arise due to the fact that the density matrix commutes with all operators. For example, by making the particle-hole transformation $\hat{c}_j \leftrightarrow \hat{c}_j^\dag$, the Hamiltonian $\hat{H}_F \rightarrow \text{const.} -\hat{H}_F$,
and we can show that
\begin{equation}
F_{jk}(t;\theta) =\frac{1}{2^L} \text{Tr}\left[  \hat{a}_j^\dag(t) \hat{a}_k^\dag(0) \hat{a}_j(t) \hat{a}_k(0) \right]
= F_{jk}(-t;\theta).
\end{equation}
Combining this with Eq.(\ref{eq:OTOC:inversion:2}) and the complex conjugation property
$[F_{jk}(t;\theta)]^\ast=F_{kj}(-t;\theta)$, we have
\begin{equation}\label{eq:OTOC:inversion:3}
  \left[F_{jk}(t;\theta) \right]^\ast =F_{kj}(-t;\theta) =F_{kj}(t;\theta)=F_{j'k'}(t;\theta).
\end{equation}
When $k=k'$ is fixed at the middle of the chain, the above equation tells us that $|F_{jk}(t)|$ is spatially symmetric as a function of the site $j$, as shown in Fig.\ref{fig:OTOC}(c).

\section{Conclusions and Discussion}\label{sec:conclusion}
We have analyzed the dynamical properties of HCAs in one-dimensional lattices by developing a general method to calculate any $n$-point dynamical correlation functions of HCAs. Our method is valid for any temperature, any statistical angle and any type of trap potentials. We have used this method to give explicit expressions of the lesser and greater Green's functions and the OTOCs. We find three main singularity lines in the spectral functions and give their dispersion relations, which can be considered as a double spectrum. These lines cross at a hot point, which induces a peak in the momentum distribution function and a power-law divergence singularity in the local spectral function. The momentum position of this hot point linearly depends on the filling factor $\nu$ and statistical angle $\theta$, and hence can be taken as an experimental signature to probe $\theta$. We also show that the anyonic statistics can induce spatial asymmetry in the Green's function, its spectral function and OTOC. In addition, the OTOCs display light-cone dynamics which is asymmetric at low temperatures but symmetric at infinite temperature.

Our results provide a way to study the dynamics of anyons in terms of the non-equilibrium Green's functions and an exact quantitative signature to probe the anyonic statistics.
Our method can be extended to calculate any $n$-point dynamical correlation functions and provide useful information for relevant experiments. It is exact at all energy and momentum scales, and hence can also be used to benchmark other approximate or phenomenological theories, such as the nonlinear Luttinger liquid theory \cite{science323-228,RevModPhys.84.1253}.
We hope this study could motivate future investigations of non-equilibrium dynamical properties of Abelian anyons in atomic, photonic and condensed matter systems.

\begin{acknowledgments}
This work has been supported by the Fundamental Research Funds for the Provincial Universities of Zhejiang,  Grant No.2021J014. We also acknowledge financial support from the Key Laboratory of Oceanographic Big Data Mining \& Application of Zhejiang Province, Zhejiang Ocean University, Zhoushan, Zhejiang, China.
\end{acknowledgments}

\appendix
\section{Analytic Properties of the Green's Functions}\label{sec:GF:analytic}
The Green's functions have the following analytic properties
\begin{eqnarray}
G^{<}_{jk}(t) &=& \frac{-i}{Z} \sum_{m,n} \langle \Phi_n|\hat{a}_j|\Phi_m\rangle \langle \Phi_m|\hat{a}_k^\dag |\Phi_n\rangle \notag\\
&&\qquad \times e^{-\beta E_m} e^{-i(E_m-E_n)t}, \\
G^{>}_{jk}(t) &=& \frac{-i}{Z} \sum_{m,n}  \langle \Phi_n|\hat{a}_j|\Phi_m\rangle \langle \Phi_m|\hat{a}_k^\dag |\Phi_n\rangle \notag\\
&& \qquad \times e^{-\beta E_n} e^{-i(E_m-E_n)t},
\end{eqnarray}
where $\{|\Phi_m\rangle \}$ are the eigenstates of the many-body Hamiltonian $\hat{H}$. Using these analytic properties it's easy to show that the spectral functions defined by Eq.(\ref{eq:spectral:real-space}) satisfies the relation $A^{-}_{jk}(\omega)=e^{-\beta\omega} A^{+}_{jk}(\omega)$ and the sum rules
\begin{equation}
 \int_{-\infty}^\infty A_{jk}^{+}(\omega)d\omega= \langle \hat{a}_j\hat{a}_k^\dag \rangle, \quad
 \int_{-\infty}^\infty A_{jk}^{-}(\omega)d\omega= \langle \hat{a}_k^\dag  \hat{a}_j \rangle.
\end{equation}
Now we prove that both the local spectral functions $A^{\pm}_{jj}(\omega)$ in real space and the momentum-space spectral functions $A^{\pm}(q,\omega)$ are nonnegative real numbers.
From the analytic properties we have
\begin{eqnarray*}
 A^{+}_{jj}(\omega) &=& \frac{1}{Z} \sum_{m,n} \langle \Phi_n|\hat{a}_j|\Phi_m\rangle \langle \Phi_m|\hat{a}_j^\dag |\Phi_n\rangle \\
 && \qquad \times e^{-\beta E_n} \delta[\omega-(E_m-E_n)] \\
 &=& \frac{1}{Z} \sum_{m,n} |\langle \Phi_n|\hat{a}_j|\Phi_m\rangle|^2  e^{-\beta E_n} \delta[\omega-(E_m-E_n)].
\end{eqnarray*}
Obviously $[A^{+}_{jj}(\omega)]^\ast=A^{+}_{jj}(\omega)\ge0$. Similarly we can prove that $[A^{-}_{jj}(\omega)]^\ast=A^{-}_{jj}(\omega)\ge0$. On the other hand, the momentum-space spectral function
\begin{eqnarray*}
 A^{+}(q,\omega) &=& \frac{1}{Z} \sum_{m,n} \frac{1}{L}\sum_{jk}e^{-iq(j-k)} \langle \Phi_n|\hat{a}_j|\Phi_m\rangle \langle \Phi_m|\hat{a}_k^\dag |\Phi_n\rangle \\
 && \qquad \times e^{-\beta E_n} \delta[\omega-(E_m-E_n)].
\end{eqnarray*}
Define operators
\begin{equation*}
  \tilde{a}_q\equiv \frac{1}{\sqrt{L}}\sum_j e^{-iqj}\hat{a}_j, \quad
  \tilde{a}_q^\dag\equiv \frac{1}{\sqrt{L}}\sum_k e^{iqk}\hat{a}_k^\dag,
\end{equation*}
then
\begin{eqnarray*}
 A^{+}(q,\omega) &=& \frac{1}{Z} \sum_{m,n}   \langle \Phi_n|\tilde{a}_q|\Phi_m\rangle \langle \Phi_m|\tilde{a}_q^\dag |\Phi_n\rangle \\
 && \qquad \times e^{-\beta E_n} \delta[\omega-(E_m-E_n)]\\
 &=& \frac{1}{Z} \sum_{m,n}   |\langle \Phi_n|\tilde{a}_q|\Phi_m\rangle|^2 e^{-\beta E_n} \delta[\omega-(E_m-E_n)].
\end{eqnarray*}
Obviously, $[A^{+}(q,\omega)]^\ast=A^{+}(q,\omega)\ge0$. Similarly we can prove that $[A^{-}(q,\omega)]^\ast=A^{-}(q,\omega)\ge0$.

\begin{widetext}
\section{Derivation of the Green's Function}\label{sec:GF:append}
We will denote $\hat{c}^\dag=(\hat{c}_1^\dag, \hat{c}_2^\dag \ldots,\hat{c}_L^\dag)$ and $\hat{c}=(\hat{c}_1, \hat{c}_2 \ldots,\hat{c}_L)^T$. A general bilinear form of the fermion operators can be written as $\hat{c}^\dag \mathds{J} \hat{c}$, where $\mathds{J}$ is a $L\times L$ matrix. The matrix $\mathds{J}$ is usually Hermitian or anti-Hermitian, but this is not necessary in general. A \emph{Gaussian operator} is defined as an operator of the form $e^{\hat{c}^\dag \mathds{J} \hat{c}}$. Two important properties of such operators are:
\begin{equation}
  e^{\hat{c}^\dag \mathds{J}_1 \hat{c}} e^{\hat{c}^\dag \mathds{J}_2 \hat{c}} =e^{\hat{c}^\dag \mathds{J} \hat{c}},
  \quad \text{with} \quad e^{\mathds{J}_1} e^{\mathds{J}_2}=e^{\mathds{J}}.
\end{equation}
\begin{equation}
  e^{\hat{c}^\dag \mathds{J} \hat{c}} \,\hat{c}\, e^{-\hat{c}^\dag \mathds{J} \hat{c}} = e^{-\mathds{J}}\hat{c}, \quad
  e^{\hat{c}^\dag \mathds{J} \hat{c}} \,\hat{c}^\dag\, e^{-\hat{c}^\dag \mathds{J} \hat{c}} = \hat{c}^\dag e^{\mathds{J}}.
\end{equation}

Now let's define a series of diagonal $L\times L$ matrices $\mathds{J}^{(m)}, m=1,2,\cdots,L$, whose diagonal matrix elements $\mathds{J}^{(m)}_{ll}$ equals to $i\theta$ for $l<m$ and 0 for $l\ge m$.
\begin{eqnarray*}
\langle \hat{a}_j(t) \hat{a}_k^\dag\rangle
&=& \text{Tr}\left[e^{i\hat{H}t} \hat{a}_j e^{-i\hat{H}t} \hat{a}_k^\dag \rho_0\right] \\
&=& \frac{1}{Z} \text{Tr}\left[e^{i\hat{H}t} e^{-\hat{c}^\dag \mathds{J}^{(j)}\hat{c}}\, \hat{c}_j e^{-i\hat{H}t} \hat{c}_k^\dag  e^{\hat{c}^\dag \mathds{J}^{(k)}\hat{c}} e^{-\beta\hat{H}} \right] \\
&=& \frac{1}{Z} \text{Tr}\left[  \hat{c}_j e^{-\hat{c}^\dag \mathds{J}^{(j)}\hat{c}}\, e^{-i\hat{H}t}  e^{\hat{c}^\dag \mathds{J}^{(k)}\hat{c}} \, \hat{c}_k^\dag e^{-\beta\hat{H}} e^{i\hat{H}t} \right] \\
&=& \frac{1}{Z} \sum_{l}  \left(e^{-\mathds{J}^{(j)}} e^{-it\mathds{H}} e^{\mathds{J}^{(k)}} \right)_{jl} \text{Tr}\left[  \hat{c}_l  \hat{c}_k^\dag  e^{-\beta\hat{H}} e^{i\hat{H}t} e^{-\hat{c}^\dag \mathds{J}^{(j)}\hat{c}}\, e^{-i\hat{H}t}  e^{\hat{c}^\dag \mathds{J}^{(k)}\hat{c}} \right] \\
&=&\frac{1}{Z}\left(  e^{-\mathds{J}^{(j)}} e^{-it\mathds{H}} e^{\mathds{J}^{(k)}} \frac{\det\left[\mathds{1}+ e^{-\beta\mathds{H}} e^{it\mathds{H}} e^{-\mathds{J}^{(j)}} e^{-it\mathds{H}} e^{\mathds{J}^{(k)}}  \right] } {\mathds{1}+  e^{-\beta\mathds{H}} e^{it\mathds{H}} e^{-\mathds{J}^{(j)}} e^{-it\mathds{H}} e^{\mathds{J}^{(k)}} } \right)_{jk} .
\end{eqnarray*}
After some straightforward simplification we can obtain the final result, Eq. (\ref{eq:GF:sol:greater}). Similarly we can obtain expressions for $\langle \hat{a}_k^\dag \hat{a}_j(t) \rangle$.

\section{Derivation of the OTOC}\label{sec:OTOC:append}
The OTOC takes the form
\begin{eqnarray}
F_{jk}(t) &\equiv & \langle \hat{a}_j^\dag(t) \hat{a}_k^\dag(0) \hat{a}_j(t) \hat{a}_k(0)\rangle
=\frac{1}{Z} \text{Tr}\left[  \hat{a}_j^\dag(t) \hat{a}_k^\dag(0) \hat{a}_j(t) \hat{a}_k(0) e^{-\beta\hat{H}}\right] \notag\\
&=& \frac{1}{Z} \text{Tr}\left[ e^{it\hat{H}} \hat{c}_j^\dag e^{\hat{c}^\dag\mathds{J}^{(j)}\hat{c}} e^{-it\hat{H}} \hat{c}_k^\dag e^{\hat{c}^\dag\mathds{J}^{(k)}\hat{c}} e^{it\hat{H}} e^{-\hat{c}^\dag\mathds{J}^{(j)}\hat{c}} \hat{c}_j e^{-it\hat{H}} e^{-\hat{c}^\dag\mathds{J}^{(k)}\hat{c}}\hat{c}_k e^{-\beta\hat{H}}\right].
\end{eqnarray}
To derive the final result, let's use some notations to simplify the formulas.
\begin{eqnarray*}
&\mathds{M}_{\pm}^{j}\equiv e^{\pm\mathds{J}^{(j)}}, & \hat{M}_{\pm}^{j}\equiv e^{\pm\hat{c}^\dag\mathds{J}^{(j)}\hat{c}}, \\
&\mathds{M}_{\pm}^{t}\equiv e^{\pm it\mathds{H}}, & \hat{M}_{\pm}^{t}\equiv e^{\pm it \hat{c}^\dag\mathds{H}\hat{c}},\\
&\mathds{M}_{\pm,\pm}^{j,t}\equiv e^{\pm\mathds{J}^{(j)}} e^{\pm it\mathds{H}}, & \hat{M}_{\pm,\pm}^{j,t}\equiv e^{\pm\hat{c}^\dag\mathds{J}^{(j)}\hat{c}} e^{\pm it \hat{c}^\dag\mathds{H}\hat{c}}, \\
&\mathds{M}_{\pm,\pm,\pm}^{j,t,k}\equiv e^{\pm\mathds{J}^{(j)}} e^{\pm it\mathds{H}} e^{\pm\mathds{J}^{(k)}}, & \hat{M}_{\pm,\pm,\pm}^{j,t,k}\equiv e^{\pm\hat{c}^\dag\mathds{J}^{(j)}\hat{c}} e^{\pm it \hat{c}^\dag\mathds{H}\hat{c}} e^{\pm\hat{c}^\dag\mathds{J}^{(k)}\hat{c}} , \\
&\cdots, & \cdots
\end{eqnarray*}
Obviously these matrices and operators are all \emph{unitary}. Then
\begin{eqnarray*}
F_{jk}(t)
&=& \frac{1}{Z} \text{Tr}\left[ e^{it\hat{H}} \hat{c}_j^\dag e^{\hat{c}^\dag\mathds{J}^{(j)}\hat{c}} e^{-it\hat{H}} \hat{c}_k^\dag e^{\hat{c}^\dag\mathds{J}^{(k)}\hat{c}} e^{it\hat{H}} e^{-\hat{c}^\dag\mathds{J}^{(j)}\hat{c}} \hat{c}_j e^{-it\hat{H}} e^{-\hat{c}^\dag\mathds{J}^{(k)}\hat{c}}\hat{c}_k e^{-\beta\hat{H}}\right] \\
&=&  \frac{1}{Z} \text{Tr}\left[ \hat{c}_l^\dag \left(\hat{M}_{+,+,-}^{t,j,t}\right) \hat{c}_k^\dag \left(\hat{M}_{+,+,-}^{k,t,j}\right) \hat{c}_j \left(\hat{M}_{-,-}^{t,k}\right) \hat{c}_k e^{-\beta\hat{H}}\right] \left(\mathds{M}^{t}_{+}\right)_{lj}\\
&=&  \frac{1}{Z} \text{Tr}\left[ \hat{c}_l^\dag  \hat{c}_m^\dag \left(\hat{M}_{+,+,-,+,+,-}^{t,j,t,k,t,j}\right) \hat{c}_j \left(\hat{M}_{-,-}^{t,k}\right) \hat{c}_k e^{-\beta\hat{H}}\right] \left(\mathds{M}_{+,+,-}^{t,j,t}\right)_{mk} \left(\mathds{M}^{t}_{+}\right)_{lj}\\
&=&\frac{1}{Z} \text{Tr}\left[ \hat{c}_l^\dag  \hat{c}_m^\dag \left(\hat{M}_{+,+,-,+,+,-}^{t,j,t,k,t,j}\right) \hat{c}_j \left(\hat{M}_{-,-}^{t,k}\right) e^{-\beta\hat{H}} \hat{c}_n \right] \left(\mathds{M}_{+,+,-}^{t,j,t}\right)_{mk} \left(\mathds{M}^{t}_{+}\right)_{lj} \left( e^{-\beta\mathds{H}} \right)_{kn} \\
&=&\frac{1}{Z} \text{Tr}\left[ \hat{c}_l^\dag  \hat{c}_m^\dag \left(\hat{M}_{+,+,-,+,+,-,-,-}^{t,j,t,k,t,j,t,k} \right)  e^{-\beta\hat{H}}\, \hat{c}_r   \hat{c}_n \right] \left(\mathds{M}_{+,+,-}^{t,j,t}\right)_{mk} \left(\mathds{M}^{t}_{+}\right)_{lj} \left( e^{-\beta\mathds{H}} \right)_{kn} \left(\mathds{M}_{-,-}^{t,k}\, e^{-\beta\mathds{H}} \right)_{jr}\\
&=& \frac{1}{Z} \text{Tr}\left[ \left(\hat{M}_{+,+,-,+,+,-,-,-}^{t,j,t,k,t,j,t,k} \right)  e^{-\beta\hat{H}}\, \hat{c}_r \hat{c}_n \hat{c}_l^\dag  \hat{c}_m^\dag \right] \left(\mathds{M}_{+,+,-}^{t,j,t}\right)_{mk} \left(\mathds{M}^{t}_{+}\right)_{lj} \left(e^{-\beta\mathds{H}} \right)_{kn} \left(\mathds{M}_{-,-}^{t,k}\, e^{-\beta\mathds{H}} \right)_{jr},
\end{eqnarray*}
where the Einstein's summation rule has been used for indices $l,m,r,n$.
Then using Wick's theorem we have
\begin{eqnarray*}
&& \frac{1}{Z} \text{Tr}\left[ \left(\hat{M}_{+,+,-,+,+,-,-,-}^{t,j,t,k,t,j,t,k} \right)  e^{-\beta\hat{H}}\, \hat{c}_r \hat{c}_n \hat{c}_l^\dag  \hat{c}_m^\dag \right]\\
&=& \frac{\det\left[\mathds{1}+ \left(\mathds{M}_{+,+,-,+,+,-,-,-}^{t,j,t,k,t,j,t,k} \right) e^{-\beta\mathds{H}}\right]}{\det\left[\mathds{1}+e^{-\beta\mathds{H}} \right]}
\left[\langle \hat{c}_{r}\hat{c}_{m}^\dag\rangle \langle \hat{c}_{n}\hat{c}_{l}^\dag\rangle -\langle \hat{c}_{r}\hat{c}_{l}^\dag\rangle \langle \hat{c}_{n}\hat{c}_{m}^\dag\rangle \right] \\
&=& \frac{\det\left[\mathds{1}+  \mathds{M}e^{-\beta\mathds{H}}\right]} {\det\left[\mathds{1}+e^{-\beta\mathds{H}} \right]} \left\{ \left[\mathds{1}+\mathds{M}e^{-\beta\mathds{H}}\right]^{-1}_{rm} \left[\mathds{1}+\mathds{M}e^{-\beta\mathds{H}}\right]^{-1}_{nl} -\left[\mathds{1}+\mathds{M}e^{-\beta\mathds{H}}\right]^{-1}_{rl} \left[\mathds{1}+\mathds{M}e^{-\beta\mathds{H}}\right]^{-1}_{nm} \right\}
\end{eqnarray*}
where $\mathds{M}\equiv \mathds{M}_{+,+,-,+,+,-,-,-}^{t,j,t,k,t,j,t,k}$. The final expression Eq.(\ref{eq:otoc:sol}) can be obtained after some algebra.

\section{From Lattice to Continuous Space}\label{sec:lattice:continuous}
The formulas for the Green's function on a \emph{lattice} can also be used to obtain the Green's function in continuous space.
Suppose that in continuous space the one-particle eigenfunctions are $\phi_n(x)=\langle x|\phi_n\rangle, n=1,2,3,\cdots$, with corresponding eigenenergies $\epsilon_n$. In ground state only the lowest $N$ levels are occupied, where $N$ is the particle number. In the above expressions we should replace the lattice sites $j$ and $k$ by corresponding spatial coordinates $x_j$ and $x_k$. The diagonal matrix $\mathds{J}^{(k)}$ should be replaced by a function of $x$ in the coordinate representation.
We have
\begin{equation}
  \mathds{B}_0=\sum_{n>N} |\phi_n\rangle\langle \phi_n|, \qquad
  \mathds{1}-\mathds{B}_0=\sum_{n=1}^N |\phi_n\rangle\langle \phi_n|.
\end{equation}
\begin{equation}
  \left(e^{\mathds{J}^{(k)}}\right)(x)=1+(e^{i\theta}-1)\Theta(x_k-x), \quad
  \left(e^{-\mathds{J}^{(k)}}\right)(x)=1+(e^{-i\theta}-1)\Theta(x_k-x),
\end{equation}
where $\Theta(x)$ is the step function. The matrix elements of $\mathds{P}^{j}_{-}(t)$ in the energy-representation are
\begin{eqnarray}
\langle \phi_m|\mathds{P}^{j}_{\pm}(t)|\phi_n\rangle
&=& e^{i(\epsilon_m-\epsilon_n)t} \langle \phi_m| e^{\pm \mathds{J}^{(j)}} |\phi_n\rangle  \notag\\
&=& e^{i(\epsilon_m-\epsilon_n)t} \int_{-\infty}^{\infty} dz\, \phi_m^\ast(z) \phi_n(z) \left(e^{\pm\mathds{J}^{(j)}}\right)(z) \notag\\
&=& \delta_{mn} + (e^{\pm i\theta}-1) e^{i(\epsilon_m-\epsilon_n)t} \int_{-\infty}^{x_j} dz\, \phi_m^\ast(z) \phi_n(z)  \notag\\
&=& e^{\pm i\theta}\delta_{mn} - (e^{\pm i\theta}-1) e^{i(\epsilon_m-\epsilon_n)t} \int_{x_j}^{\infty} dz\, \phi_m^\ast(z) \phi_n(z).
\end{eqnarray}

It's convenient to separate the total Hilbert space $\mathcal{H}$ into two subspaces, $\mathcal{H}=\mathcal{H}_a\bigoplus\mathcal{H}_b$, where
\begin{equation*}
\mathcal{H}_a =\text{span}\left\{|\phi_1\rangle, \cdots, |\phi_N\rangle\right\},
\quad \mathcal{H}_b =\text{span}\left\{|\phi_{N+1}\rangle, |\phi_{N+2}\rangle,\cdots \right\}.
\end{equation*}
Then any matrix (operator) $\mathds{A}$ can be expressed in a block form,
\begin{equation*}
  \mathds{A}=\left(
               \begin{array}{cc}
                 \mathds{A}_{aa} & \mathds{A}_{ab} \\
                 \mathds{A}_{ba} & \mathds{A}_{bb} \\
               \end{array}
             \right).
\end{equation*}
For example,
\begin{equation*}
  \mathds{B}_0= \left(
               \begin{array}{cc}
                 0 & 0 \\
                 0 & \mathds{1}_{bb} \\
               \end{array}
             \right), \quad
  \mathds{1}-\mathds{B}_0= \left(
               \begin{array}{cc}
                 \mathds{1}_{aa} & 0 \\
                 0 & 0 \\
               \end{array}
             \right).
\end{equation*}
Furthermore,
\begin{equation*}
  \mathds{B}^{jk}(t)= \left(
               \begin{array}{cc}
               [\mathds{P}^{j}_{-}(t)\mathds{P}^{k}_{+}(0)]_{aa} & 0 \\
               {}[\mathds{P}^{j}_{-}(t)\mathds{P}^{k}_{+}(0)]_{ba} & \mathds{1}_{bb} \\
               \end{array}
             \right), \quad
  (\mathds{1}-\mathds{B}_0)\left[\mathds{B}^{jk}(t)\right]^{-1} =\left(
               \begin{array}{cc}
                 \left\{[\mathds{P}^{j}_{-}(t)\mathds{P}^{k}_{+}(0)]_{aa}\right\}^{-1} & 0 \\
                 0 & 0 \\
               \end{array}
             \right).
\end{equation*}
Therefore,
\begin{equation}
  \det[\mathds{B}^{jk}(t)] = \det\left\{[\mathds{P}^{j}_{-}(t)\mathds{P}^{k}_{+}(0)]_{aa}\right\},
\end{equation}
i.e., the determinant of the infinite matrix $\mathds{B}^{jk}(t)$ can  be expressed as a determinant of a finite $N\times N$ matrix.

Let's denote $\Phi(x)=(\phi_1(x), \phi_2(x),\ldots)^T$ as the column vector of the single-particle orbitals, and $\Phi(x,t)=(e^{-i\epsilon_1 t}\phi_1(x), e^{-i\epsilon_2 t}\phi_2(x),\ldots)^T$.
Then
\begin{eqnarray*}
\left\{ e^{-it\mathds{H}} \right\}_{jk} &=& \Phi(x_j,t)^T \Phi(x_k)^\ast =\Phi(x_k)^\dag \Phi(x_j,t), \\
\left\{ e^{-it\mathds{H}} \mathds{P}^{k}_{+}(0) (\mathds{1}-\mathds{B}_0)\left[\mathds{B}^{jk}(t)\right]^{-1} \mathds{P}^{j}_{-}(t) \right\}_{jk}
&=&\sum_{m,n=1}^{N} \left[\Phi(x_j,t)^T \mathds{P}^{k}_{+}(0) \right]_{m} \left\{[\mathds{P}^{j}_{-}(t)\mathds{P}^{k}_{+}(0)]_{aa}\right\}^{-1}_{mn} \left[\mathds{P}^{j}_{-}(t) \Phi(x_k)^\ast\right]_n.
\end{eqnarray*}
Then the greater Green's function reads
\begin{equation}
iG^{>}_{jk}(t)= \langle \hat{a}_j(t) \hat{a}_k^\dag\rangle
=\det\left\{[\mathds{P}^{j}_{-}(t)\mathds{P}^{k}_{+}(0)]_{aa}\right\}  a^{>}(x_j,x_k,t),
\end{equation}
where
\begin{equation}
a^{>}(x_j,x_k,t) = \Phi(x_j,t)^T \Phi(x_k)^\ast
 -\sum_{m,n=1}^{N} \left[\Phi(x_j,t)^T \mathds{P}^{k}_{+}(0) \right]_{m} \left\{[\mathds{P}^{j}_{-}(t)\mathds{P}^{k}_{+}(0)]_{aa}\right\}^{-1}_{mn} \left[\mathds{P}^{j}_{-}(t) \Phi(x_k)^\ast\right]_n.
\end{equation}
This result is essentially equivalent to the expressions given in a recent Letter \cite{PhysRevLett.126.065301}. Similarly we can extend the expression for the lesser Green's function, which is simpler, and that for the OTOC, which is more complicated, to continuous space.  We would not elaborate to give the details here.

\end{widetext}

\bibliography{anyon-hard-core-dyn-bib}

\begin{thebibliography}{94}%
\makeatletter
\providecommand \@ifxundefined [1]{%
 \@ifx{#1\undefined}
}%
\providecommand \@ifnum [1]{%
 \ifnum #1\expandafter \@firstoftwo
 \else \expandafter \@secondoftwo
 \fi
}%
\providecommand \@ifx [1]{%
 \ifx #1\expandafter \@firstoftwo
 \else \expandafter \@secondoftwo
 \fi
}%
\providecommand \natexlab [1]{#1}%
\providecommand \enquote  [1]{``#1''}%
\providecommand \bibnamefont  [1]{#1}%
\providecommand \bibfnamefont [1]{#1}%
\providecommand \citenamefont [1]{#1}%
\providecommand \href@noop [0]{\@secondoftwo}%
\providecommand \href [0]{\begingroup \@sanitize@url \@href}%
\providecommand \@href[1]{\@@startlink{#1}\@@href}%
\providecommand \@@href[1]{\endgroup#1\@@endlink}%
\providecommand \@sanitize@url [0]{\catcode `\\12\catcode `\$12\catcode
  `\&12\catcode `\#12\catcode `\^12\catcode `\_12\catcode `\%12\relax}%
\providecommand \@@startlink[1]{}%
\providecommand \@@endlink[0]{}%
\providecommand \url  [0]{\begingroup\@sanitize@url \@url }%
\providecommand \@url [1]{\endgroup\@href {#1}{\urlprefix }}%
\providecommand \urlprefix  [0]{URL }%
\providecommand \Eprint [0]{\href }%
\providecommand \doibase [0]{http://dx.doi.org/}%
\providecommand \selectlanguage [0]{\@gobble}%
\providecommand \bibinfo  [0]{\@secondoftwo}%
\providecommand \bibfield  [0]{\@secondoftwo}%
\providecommand \translation [1]{[#1]}%
\providecommand \BibitemOpen [0]{}%
\providecommand \bibitemStop [0]{}%
\providecommand \bibitemNoStop [0]{.\EOS\space}%
\providecommand \EOS [0]{\spacefactor3000\relax}%
\providecommand \BibitemShut  [1]{\csname bibitem#1\endcsname}%
\let\auto@bib@innerbib\@empty
\bibitem [{\citenamefont {Leinaas}\ and\ \citenamefont
  {Myrheim}(1977)}]{nuovo1977}%
  \BibitemOpen
  \bibfield  {author} {\bibinfo {author} {\bibfnamefont {J.}~\bibnamefont
  {Leinaas}}\ and\ \bibinfo {author} {\bibfnamefont {J.}~\bibnamefont
  {Myrheim}},\ }\href {\doibase https://doi.org/10.1007/BF02727953} {\bibfield
  {journal} {\bibinfo  {journal} {Nuovo Cimento Soc. Ital. Fis. B}\ }\textbf
  {\bibinfo {volume} {37}},\ \bibinfo {pages} {1} (\bibinfo {year}
  {1977})}\BibitemShut {NoStop}%
\bibitem [{\citenamefont {Goldin}\ \emph {et~al.}(1981)\citenamefont {Goldin},
  \citenamefont {Menikoff},\ and\ \citenamefont {Sharp}}]{JMP22-1664}%
  \BibitemOpen
  \bibfield  {author} {\bibinfo {author} {\bibfnamefont {G.~A.}\ \bibnamefont
  {Goldin}}, \bibinfo {author} {\bibfnamefont {R.}~\bibnamefont {Menikoff}}, \
  and\ \bibinfo {author} {\bibfnamefont {D.~H.}\ \bibnamefont {Sharp}},\ }\href
  {\doibase 10.1063/1.525110} {\bibfield  {journal} {\bibinfo  {journal} {J.
  Math. Phys.}\ }\textbf {\bibinfo {volume} {22}},\ \bibinfo {pages} {1664}
  (\bibinfo {year} {1981})}\BibitemShut {NoStop}%
\bibitem [{\citenamefont {Wilczek}(1982)}]{PhysRevLett.48.1144}%
  \BibitemOpen
  \bibfield  {author} {\bibinfo {author} {\bibfnamefont {F.}~\bibnamefont
  {Wilczek}},\ }\href {\doibase 10.1103/PhysRevLett.48.1144} {\bibfield
  {journal} {\bibinfo  {journal} {Phys. Rev. Lett.}\ }\textbf {\bibinfo
  {volume} {48}},\ \bibinfo {pages} {1144} (\bibinfo {year}
  {1982})}\BibitemShut {NoStop}%
\bibitem [{\citenamefont {Tsui}\ \emph {et~al.}(1982)\citenamefont {Tsui},
  \citenamefont {Stormer},\ and\ \citenamefont
  {Gossard}}]{PhysRevLett.48.1559}%
  \BibitemOpen
  \bibfield  {author} {\bibinfo {author} {\bibfnamefont {D.~C.}\ \bibnamefont
  {Tsui}}, \bibinfo {author} {\bibfnamefont {H.~L.}\ \bibnamefont {Stormer}}, \
  and\ \bibinfo {author} {\bibfnamefont {A.~C.}\ \bibnamefont {Gossard}},\
  }\href {\doibase 10.1103/PhysRevLett.48.1559} {\bibfield  {journal} {\bibinfo
   {journal} {Phys. Rev. Lett.}\ }\textbf {\bibinfo {volume} {48}},\ \bibinfo
  {pages} {1559} (\bibinfo {year} {1982})}\BibitemShut {NoStop}%
\bibitem [{\citenamefont {Laughlin}(1983)}]{PhysRevLett.50.1395}%
  \BibitemOpen
  \bibfield  {author} {\bibinfo {author} {\bibfnamefont {R.~B.}\ \bibnamefont
  {Laughlin}},\ }\href {\doibase 10.1103/PhysRevLett.50.1395} {\bibfield
  {journal} {\bibinfo  {journal} {Phys. Rev. Lett.}\ }\textbf {\bibinfo
  {volume} {50}},\ \bibinfo {pages} {1395} (\bibinfo {year}
  {1983})}\BibitemShut {NoStop}%
\bibitem [{\citenamefont {Halperin}(1984)}]{PhysRevLett.52.1583}%
  \BibitemOpen
  \bibfield  {author} {\bibinfo {author} {\bibfnamefont {B.~I.}\ \bibnamefont
  {Halperin}},\ }\href {\doibase 10.1103/PhysRevLett.52.1583} {\bibfield
  {journal} {\bibinfo  {journal} {Phys. Rev. Lett.}\ }\textbf {\bibinfo
  {volume} {52}},\ \bibinfo {pages} {1583} (\bibinfo {year}
  {1984})}\BibitemShut {NoStop}%
\bibitem [{\citenamefont {Arovas}\ \emph {et~al.}(1984)\citenamefont {Arovas},
  \citenamefont {Schrieffer},\ and\ \citenamefont
  {Wilczek}}]{PhysRevLett.53.722}%
  \BibitemOpen
  \bibfield  {author} {\bibinfo {author} {\bibfnamefont {D.}~\bibnamefont
  {Arovas}}, \bibinfo {author} {\bibfnamefont {J.~R.}\ \bibnamefont
  {Schrieffer}}, \ and\ \bibinfo {author} {\bibfnamefont {F.}~\bibnamefont
  {Wilczek}},\ }\href {\doibase 10.1103/PhysRevLett.53.722} {\bibfield
  {journal} {\bibinfo  {journal} {Phys. Rev. Lett.}\ }\textbf {\bibinfo
  {volume} {53}},\ \bibinfo {pages} {722} (\bibinfo {year} {1984})}\BibitemShut
  {NoStop}%
\bibitem [{\citenamefont {Kitaev}(2003)}]{KITAEV2003}%
  \BibitemOpen
  \bibfield  {author} {\bibinfo {author} {\bibfnamefont {A.}~\bibnamefont
  {Kitaev}},\ }\href@noop {} {\bibfield  {journal} {\bibinfo  {journal} {Ann.
  Phys.}\ }\textbf {\bibinfo {volume} {303}},\ \bibinfo {pages} {2} (\bibinfo
  {year} {2003})}\BibitemShut {NoStop}%
\bibitem [{\citenamefont {Das~Sarma}\ \emph {et~al.}(2005)\citenamefont
  {Das~Sarma}, \citenamefont {Freedman},\ and\ \citenamefont
  {Nayak}}]{PhysRevLett.94.166802}%
  \BibitemOpen
  \bibfield  {author} {\bibinfo {author} {\bibfnamefont {S.}~\bibnamefont
  {Das~Sarma}}, \bibinfo {author} {\bibfnamefont {M.}~\bibnamefont {Freedman}},
  \ and\ \bibinfo {author} {\bibfnamefont {C.}~\bibnamefont {Nayak}},\ }\href
  {\doibase 10.1103/PhysRevLett.94.166802} {\bibfield  {journal} {\bibinfo
  {journal} {Phys. Rev. Lett.}\ }\textbf {\bibinfo {volume} {94}},\ \bibinfo
  {pages} {166802} (\bibinfo {year} {2005})}\BibitemShut {NoStop}%
\bibitem [{\citenamefont {Nayak}\ \emph {et~al.}(2008)\citenamefont {Nayak},
  \citenamefont {Simon}, \citenamefont {Stern}, \citenamefont {Freedman},\ and\
  \citenamefont {Das~Sarma}}]{RevModPhys.80.1083}%
  \BibitemOpen
  \bibfield  {author} {\bibinfo {author} {\bibfnamefont {C.}~\bibnamefont
  {Nayak}}, \bibinfo {author} {\bibfnamefont {S.~H.}\ \bibnamefont {Simon}},
  \bibinfo {author} {\bibfnamefont {A.}~\bibnamefont {Stern}}, \bibinfo
  {author} {\bibfnamefont {M.}~\bibnamefont {Freedman}}, \ and\ \bibinfo
  {author} {\bibfnamefont {S.}~\bibnamefont {Das~Sarma}},\ }\href {\doibase
  10.1103/RevModPhys.80.1083} {\bibfield  {journal} {\bibinfo  {journal} {Rev.
  Mod. Phys.}\ }\textbf {\bibinfo {volume} {80}},\ \bibinfo {pages} {1083}
  (\bibinfo {year} {2008})}\BibitemShut {NoStop}%
\bibitem [{\citenamefont {Stern}\ and\ \citenamefont
  {Lindner}(2013)}]{science339-1179}%
  \BibitemOpen
  \bibfield  {author} {\bibinfo {author} {\bibfnamefont {A.}~\bibnamefont
  {Stern}}\ and\ \bibinfo {author} {\bibfnamefont {N.~H.}\ \bibnamefont
  {Lindner}},\ }\href {\doibase 10.1126/science.1231473} {\bibfield  {journal}
  {\bibinfo  {journal} {Science}\ }\textbf {\bibinfo {volume} {339}},\ \bibinfo
  {pages} {1179} (\bibinfo {year} {2013})}\BibitemShut {NoStop}%
\bibitem [{\citenamefont {Kitaev}(2006)}]{KITAEV20062}%
  \BibitemOpen
  \bibfield  {author} {\bibinfo {author} {\bibfnamefont {A.}~\bibnamefont
  {Kitaev}},\ }\href {\doibase https://doi.org/10.1016/j.aop.2005.10.005}
  {\bibfield  {journal} {\bibinfo  {journal} {Ann. Phys.}\ }\textbf {\bibinfo
  {volume} {321}},\ \bibinfo {pages} {2} (\bibinfo {year} {2006})}\BibitemShut
  {NoStop}%
\bibitem [{\citenamefont {Yao}\ and\ \citenamefont
  {Kivelson}(2007)}]{PhysRevLett.99.247203}%
  \BibitemOpen
  \bibfield  {author} {\bibinfo {author} {\bibfnamefont {H.}~\bibnamefont
  {Yao}}\ and\ \bibinfo {author} {\bibfnamefont {S.~A.}\ \bibnamefont
  {Kivelson}},\ }\href {\doibase 10.1103/PhysRevLett.99.247203} {\bibfield
  {journal} {\bibinfo  {journal} {Phys. Rev. Lett.}\ }\textbf {\bibinfo
  {volume} {99}},\ \bibinfo {pages} {247203} (\bibinfo {year}
  {2007})}\BibitemShut {NoStop}%
\bibitem [{\citenamefont {Haldane}(1991{\natexlab{a}})}]{PhysRevLett.67.937}%
  \BibitemOpen
  \bibfield  {author} {\bibinfo {author} {\bibfnamefont {F.~D.~M.}\
  \bibnamefont {Haldane}},\ }\href {\doibase 10.1103/PhysRevLett.67.937}
  {\bibfield  {journal} {\bibinfo  {journal} {Phys. Rev. Lett.}\ }\textbf
  {\bibinfo {volume} {67}},\ \bibinfo {pages} {937} (\bibinfo {year}
  {1991}{\natexlab{a}})}\BibitemShut {NoStop}%
\bibitem [{\citenamefont {Haldane}(1991{\natexlab{b}})}]{PhysRevLett.66.1529}%
  \BibitemOpen
  \bibfield  {author} {\bibinfo {author} {\bibfnamefont {F.~D.~M.}\
  \bibnamefont {Haldane}},\ }\href {\doibase 10.1103/PhysRevLett.66.1529}
  {\bibfield  {journal} {\bibinfo  {journal} {Phys. Rev. Lett.}\ }\textbf
  {\bibinfo {volume} {66}},\ \bibinfo {pages} {1529} (\bibinfo {year}
  {1991}{\natexlab{b}})}\BibitemShut {NoStop}%
\bibitem [{\citenamefont {Ha}(1994)}]{PhysRevLett.73.1574}%
  \BibitemOpen
  \bibfield  {author} {\bibinfo {author} {\bibfnamefont {Z.~N.~C.}\
  \bibnamefont {Ha}},\ }\href {\doibase 10.1103/PhysRevLett.73.1574} {\bibfield
   {journal} {\bibinfo  {journal} {Phys. Rev. Lett.}\ }\textbf {\bibinfo
  {volume} {73}},\ \bibinfo {pages} {1574} (\bibinfo {year}
  {1994})}\BibitemShut {NoStop}%
\bibitem [{\citenamefont {Murthy}\ and\ \citenamefont
  {Shankar}(1994)}]{PhysRevLett.73.3331}%
  \BibitemOpen
  \bibfield  {author} {\bibinfo {author} {\bibfnamefont {M.~V.~N.}\
  \bibnamefont {Murthy}}\ and\ \bibinfo {author} {\bibfnamefont
  {R.}~\bibnamefont {Shankar}},\ }\href {\doibase 10.1103/PhysRevLett.73.3331}
  {\bibfield  {journal} {\bibinfo  {journal} {Phys. Rev. Lett.}\ }\textbf
  {\bibinfo {volume} {73}},\ \bibinfo {pages} {3331} (\bibinfo {year}
  {1994})}\BibitemShut {NoStop}%
\bibitem [{\citenamefont {Wu}\ and\ \citenamefont
  {Yu}(1995)}]{PhysRevLett.75.890}%
  \BibitemOpen
  \bibfield  {author} {\bibinfo {author} {\bibfnamefont {Y.-S.}\ \bibnamefont
  {Wu}}\ and\ \bibinfo {author} {\bibfnamefont {Y.}~\bibnamefont {Yu}},\ }\href
  {\doibase 10.1103/PhysRevLett.75.890} {\bibfield  {journal} {\bibinfo
  {journal} {Phys. Rev. Lett.}\ }\textbf {\bibinfo {volume} {75}},\ \bibinfo
  {pages} {890} (\bibinfo {year} {1995})}\BibitemShut {NoStop}%
\bibitem [{\citenamefont {Amico}\ \emph {et~al.}(1998)\citenamefont {Amico},
  \citenamefont {Osterloh},\ and\ \citenamefont {Eckern}}]{PhysRevB.58.R1703}%
  \BibitemOpen
  \bibfield  {author} {\bibinfo {author} {\bibfnamefont {L.}~\bibnamefont
  {Amico}}, \bibinfo {author} {\bibfnamefont {A.}~\bibnamefont {Osterloh}}, \
  and\ \bibinfo {author} {\bibfnamefont {U.}~\bibnamefont {Eckern}},\ }\href
  {\doibase 10.1103/PhysRevB.58.R1703} {\bibfield  {journal} {\bibinfo
  {journal} {Phys. Rev. B}\ }\textbf {\bibinfo {volume} {58}},\ \bibinfo
  {pages} {R1703} (\bibinfo {year} {1998})}\BibitemShut {NoStop}%
\bibitem [{\citenamefont {Mazza}\ \emph {et~al.}(2018)\citenamefont {Mazza},
  \citenamefont {Viti}, \citenamefont {Carrega}, \citenamefont {Rossini},\ and\
  \citenamefont {De~Luca}}]{PhysRevB.98.075421}%
  \BibitemOpen
  \bibfield  {author} {\bibinfo {author} {\bibfnamefont {L.}~\bibnamefont
  {Mazza}}, \bibinfo {author} {\bibfnamefont {J.}~\bibnamefont {Viti}},
  \bibinfo {author} {\bibfnamefont {M.}~\bibnamefont {Carrega}}, \bibinfo
  {author} {\bibfnamefont {D.}~\bibnamefont {Rossini}}, \ and\ \bibinfo
  {author} {\bibfnamefont {A.}~\bibnamefont {De~Luca}},\ }\href {\doibase
  10.1103/PhysRevB.98.075421} {\bibfield  {journal} {\bibinfo  {journal} {Phys.
  Rev. B}\ }\textbf {\bibinfo {volume} {98}},\ \bibinfo {pages} {075421}
  (\bibinfo {year} {2018})}\BibitemShut {NoStop}%
\bibitem [{\citenamefont {Zinner}(2015)}]{PhysRevA.92.063634}%
  \BibitemOpen
  \bibfield  {author} {\bibinfo {author} {\bibfnamefont {N.~T.}\ \bibnamefont
  {Zinner}},\ }\href {\doibase 10.1103/PhysRevA.92.063634} {\bibfield
  {journal} {\bibinfo  {journal} {Phys. Rev. A}\ }\textbf {\bibinfo {volume}
  {92}},\ \bibinfo {pages} {063634} (\bibinfo {year} {2015})}\BibitemShut
  {NoStop}%
\bibitem [{\citenamefont {Kundu}(1999)}]{PhysRevLett.83.1275}%
  \BibitemOpen
  \bibfield  {author} {\bibinfo {author} {\bibfnamefont {A.}~\bibnamefont
  {Kundu}},\ }\href {\doibase 10.1103/PhysRevLett.83.1275} {\bibfield
  {journal} {\bibinfo  {journal} {Phys. Rev. Lett.}\ }\textbf {\bibinfo
  {volume} {83}},\ \bibinfo {pages} {1275} (\bibinfo {year}
  {1999})}\BibitemShut {NoStop}%
\bibitem [{\citenamefont {Batchelor}\ \emph {et~al.}(2006)\citenamefont
  {Batchelor}, \citenamefont {Guan},\ and\ \citenamefont
  {Oelkers}}]{PhysRevLett.96.210402}%
  \BibitemOpen
  \bibfield  {author} {\bibinfo {author} {\bibfnamefont {M.~T.}\ \bibnamefont
  {Batchelor}}, \bibinfo {author} {\bibfnamefont {X.-W.}\ \bibnamefont {Guan}},
  \ and\ \bibinfo {author} {\bibfnamefont {N.}~\bibnamefont {Oelkers}},\ }\href
  {\doibase 10.1103/PhysRevLett.96.210402} {\bibfield  {journal} {\bibinfo
  {journal} {Phys. Rev. Lett.}\ }\textbf {\bibinfo {volume} {96}},\ \bibinfo
  {pages} {210402} (\bibinfo {year} {2006})}\BibitemShut {NoStop}%
\bibitem [{\citenamefont {Girardeau}(2006)}]{PhysRevLett.97.100402}%
  \BibitemOpen
  \bibfield  {author} {\bibinfo {author} {\bibfnamefont {M.~D.}\ \bibnamefont
  {Girardeau}},\ }\href {\doibase 10.1103/PhysRevLett.97.100402} {\bibfield
  {journal} {\bibinfo  {journal} {Phys. Rev. Lett.}\ }\textbf {\bibinfo
  {volume} {97}},\ \bibinfo {pages} {100402} (\bibinfo {year}
  {2006})}\BibitemShut {NoStop}%
\bibitem [{\citenamefont {Greiter}(2009)}]{PhysRevB.79.064409}%
  \BibitemOpen
  \bibfield  {author} {\bibinfo {author} {\bibfnamefont {M.}~\bibnamefont
  {Greiter}},\ }\href {\doibase 10.1103/PhysRevB.79.064409} {\bibfield
  {journal} {\bibinfo  {journal} {Phys. Rev. B}\ }\textbf {\bibinfo {volume}
  {79}},\ \bibinfo {pages} {064409} (\bibinfo {year} {2009})}\BibitemShut
  {NoStop}%
\bibitem [{\citenamefont {Calabrese}\ and\ \citenamefont
  {Mintchev}(2007)}]{PhysRevB.75.233104}%
  \BibitemOpen
  \bibfield  {author} {\bibinfo {author} {\bibfnamefont {P.}~\bibnamefont
  {Calabrese}}\ and\ \bibinfo {author} {\bibfnamefont {M.}~\bibnamefont
  {Mintchev}},\ }\href {\doibase 10.1103/PhysRevB.75.233104} {\bibfield
  {journal} {\bibinfo  {journal} {Phys. Rev. B}\ }\textbf {\bibinfo {volume}
  {75}},\ \bibinfo {pages} {233104} (\bibinfo {year} {2007})}\BibitemShut
  {NoStop}%
\bibitem [{\citenamefont {Hao}\ \emph {et~al.}(2008)\citenamefont {Hao},
  \citenamefont {Zhang},\ and\ \citenamefont {Chen}}]{PhysRevA.78.023631}%
  \BibitemOpen
  \bibfield  {author} {\bibinfo {author} {\bibfnamefont {Y.}~\bibnamefont
  {Hao}}, \bibinfo {author} {\bibfnamefont {Y.}~\bibnamefont {Zhang}}, \ and\
  \bibinfo {author} {\bibfnamefont {S.}~\bibnamefont {Chen}},\ }\href {\doibase
  10.1103/PhysRevA.78.023631} {\bibfield  {journal} {\bibinfo  {journal} {Phys.
  Rev. A}\ }\textbf {\bibinfo {volume} {78}},\ \bibinfo {pages} {023631}
  (\bibinfo {year} {2008})}\BibitemShut {NoStop}%
\bibitem [{\citenamefont {Hao}\ \emph {et~al.}(2009)\citenamefont {Hao},
  \citenamefont {Zhang},\ and\ \citenamefont {Chen}}]{PhysRevA.79.043633}%
  \BibitemOpen
  \bibfield  {author} {\bibinfo {author} {\bibfnamefont {Y.}~\bibnamefont
  {Hao}}, \bibinfo {author} {\bibfnamefont {Y.}~\bibnamefont {Zhang}}, \ and\
  \bibinfo {author} {\bibfnamefont {S.}~\bibnamefont {Chen}},\ }\href {\doibase
  10.1103/PhysRevA.79.043633} {\bibfield  {journal} {\bibinfo  {journal} {Phys.
  Rev. A}\ }\textbf {\bibinfo {volume} {79}},\ \bibinfo {pages} {043633}
  (\bibinfo {year} {2009})}\BibitemShut {NoStop}%
\bibitem [{\citenamefont {Tang}\ \emph {et~al.}(2015)\citenamefont {Tang},
  \citenamefont {Eggert},\ and\ \citenamefont {Pelster}}]{Tang_2015}%
  \BibitemOpen
  \bibfield  {author} {\bibinfo {author} {\bibfnamefont {G.}~\bibnamefont
  {Tang}}, \bibinfo {author} {\bibfnamefont {S.}~\bibnamefont {Eggert}}, \ and\
  \bibinfo {author} {\bibfnamefont {A.}~\bibnamefont {Pelster}},\ }\href
  {\doibase 10.1088/1367-2630/17/12/123016} {\bibfield  {journal} {\bibinfo
  {journal} {New J. Phys.}\ }\textbf {\bibinfo {volume} {17}},\ \bibinfo
  {pages} {123016} (\bibinfo {year} {2015})}\BibitemShut {NoStop}%
\bibitem [{\citenamefont {Zatloukal}\ \emph {et~al.}(2014)\citenamefont
  {Zatloukal}, \citenamefont {Lehman}, \citenamefont {Singh}, \citenamefont
  {Pachos},\ and\ \citenamefont {Brennen}}]{PhysRevB.90.134201}%
  \BibitemOpen
  \bibfield  {author} {\bibinfo {author} {\bibfnamefont {V.}~\bibnamefont
  {Zatloukal}}, \bibinfo {author} {\bibfnamefont {L.}~\bibnamefont {Lehman}},
  \bibinfo {author} {\bibfnamefont {S.}~\bibnamefont {Singh}}, \bibinfo
  {author} {\bibfnamefont {J.~K.}\ \bibnamefont {Pachos}}, \ and\ \bibinfo
  {author} {\bibfnamefont {G.~K.}\ \bibnamefont {Brennen}},\ }\href {\doibase
  10.1103/PhysRevB.90.134201} {\bibfield  {journal} {\bibinfo  {journal} {Phys.
  Rev. B}\ }\textbf {\bibinfo {volume} {90}},\ \bibinfo {pages} {134201}
  (\bibinfo {year} {2014})}\BibitemShut {NoStop}%
\bibitem [{\citenamefont {del Campo}(2008)}]{PhysRevA.78.045602}%
  \BibitemOpen
  \bibfield  {author} {\bibinfo {author} {\bibfnamefont {A.}~\bibnamefont {del
  Campo}},\ }\href {\doibase 10.1103/PhysRevA.78.045602} {\bibfield  {journal}
  {\bibinfo  {journal} {Phys. Rev. A}\ }\textbf {\bibinfo {volume} {78}},\
  \bibinfo {pages} {045602} (\bibinfo {year} {2008})}\BibitemShut {NoStop}%
\bibitem [{\citenamefont {Hao}\ and\ \citenamefont
  {Chen}(2012)}]{PhysRevA.86.043631}%
  \BibitemOpen
  \bibfield  {author} {\bibinfo {author} {\bibfnamefont {Y.}~\bibnamefont
  {Hao}}\ and\ \bibinfo {author} {\bibfnamefont {S.}~\bibnamefont {Chen}},\
  }\href {\doibase 10.1103/PhysRevA.86.043631} {\bibfield  {journal} {\bibinfo
  {journal} {Phys. Rev. A}\ }\textbf {\bibinfo {volume} {86}},\ \bibinfo
  {pages} {043631} (\bibinfo {year} {2012})}\BibitemShut {NoStop}%
\bibitem [{\citenamefont {Piroli}\ and\ \citenamefont
  {Calabrese}(2017)}]{PhysRevA.96.023611}%
  \BibitemOpen
  \bibfield  {author} {\bibinfo {author} {\bibfnamefont {L.}~\bibnamefont
  {Piroli}}\ and\ \bibinfo {author} {\bibfnamefont {P.}~\bibnamefont
  {Calabrese}},\ }\href {\doibase 10.1103/PhysRevA.96.023611} {\bibfield
  {journal} {\bibinfo  {journal} {Phys. Rev. A}\ }\textbf {\bibinfo {volume}
  {96}},\ \bibinfo {pages} {023611} (\bibinfo {year} {2017})}\BibitemShut
  {NoStop}%
\bibitem [{\citenamefont {Wilson}\ \emph {et~al.}(2020)\citenamefont {Wilson},
  \citenamefont {Malvania}, \citenamefont {Le}, \citenamefont {Zhang},
  \citenamefont {Rigol},\ and\ \citenamefont {Weiss}}]{science367-1461}%
  \BibitemOpen
  \bibfield  {author} {\bibinfo {author} {\bibfnamefont {J.~M.}\ \bibnamefont
  {Wilson}}, \bibinfo {author} {\bibfnamefont {N.}~\bibnamefont {Malvania}},
  \bibinfo {author} {\bibfnamefont {Y.}~\bibnamefont {Le}}, \bibinfo {author}
  {\bibfnamefont {Y.}~\bibnamefont {Zhang}}, \bibinfo {author} {\bibfnamefont
  {M.}~\bibnamefont {Rigol}}, \ and\ \bibinfo {author} {\bibfnamefont {D.~S.}\
  \bibnamefont {Weiss}},\ }\href@noop {} {\bibfield  {journal} {\bibinfo
  {journal} {Science}\ }\textbf {\bibinfo {volume} {367}},\ \bibinfo {pages}
  {1461} (\bibinfo {year} {2020})}\BibitemShut {NoStop}%
\bibitem [{\citenamefont {P{\^{a}}{\c{t}}u}\ \emph {et~al.}(2007)\citenamefont
  {P{\^{a}}{\c{t}}u}, \citenamefont {Korepin},\ and\ \citenamefont
  {Averin}}]{Pu_2007}%
  \BibitemOpen
  \bibfield  {author} {\bibinfo {author} {\bibfnamefont {O.~I.}\ \bibnamefont
  {P{\^{a}}{\c{t}}u}}, \bibinfo {author} {\bibfnamefont {V.~E.}\ \bibnamefont
  {Korepin}}, \ and\ \bibinfo {author} {\bibfnamefont {D.~V.}\ \bibnamefont
  {Averin}},\ }\href {\doibase 10.1088/1751-8113/40/50/004} {\bibfield
  {journal} {\bibinfo  {journal} {J. Phys. A}\ }\textbf {\bibinfo {volume}
  {40}},\ \bibinfo {pages} {14963} (\bibinfo {year} {2007})}\BibitemShut
  {NoStop}%
\bibitem [{\citenamefont {Calabrese}\ and\ \citenamefont
  {Santachiara}(2009)}]{Calabrese_2009}%
  \BibitemOpen
  \bibfield  {author} {\bibinfo {author} {\bibfnamefont {P.}~\bibnamefont
  {Calabrese}}\ and\ \bibinfo {author} {\bibfnamefont {R.}~\bibnamefont
  {Santachiara}},\ }\href {\doibase 10.1088/1742-5468/2009/03/p03002}
  {\bibfield  {journal} {\bibinfo  {journal} {J. Stat. Mech. Theory Exp.}\
  }\textbf {\bibinfo {volume} {2009}},\ \bibinfo {pages} {P03002} (\bibinfo
  {year} {2009})}\BibitemShut {NoStop}%
\bibitem [{\citenamefont {Lange}\ \emph
  {et~al.}(2017{\natexlab{a}})\citenamefont {Lange}, \citenamefont {Ejima},\
  and\ \citenamefont {Fehske}}]{PhysRevLett.118.120401}%
  \BibitemOpen
  \bibfield  {author} {\bibinfo {author} {\bibfnamefont {F.}~\bibnamefont
  {Lange}}, \bibinfo {author} {\bibfnamefont {S.}~\bibnamefont {Ejima}}, \ and\
  \bibinfo {author} {\bibfnamefont {H.}~\bibnamefont {Fehske}},\ }\href
  {\doibase 10.1103/PhysRevLett.118.120401} {\bibfield  {journal} {\bibinfo
  {journal} {Phys. Rev. Lett.}\ }\textbf {\bibinfo {volume} {118}},\ \bibinfo
  {pages} {120401} (\bibinfo {year} {2017}{\natexlab{a}})}\BibitemShut
  {NoStop}%
\bibitem [{\citenamefont {Keilmann}\ \emph {et~al.}(2011)\citenamefont
  {Keilmann}, \citenamefont {Lanzmich}, \citenamefont {McCulloch},\ and\
  \citenamefont {Roncaglia}}]{nc2-361}%
  \BibitemOpen
  \bibfield  {author} {\bibinfo {author} {\bibfnamefont {T.}~\bibnamefont
  {Keilmann}}, \bibinfo {author} {\bibfnamefont {S.}~\bibnamefont {Lanzmich}},
  \bibinfo {author} {\bibfnamefont {I.}~\bibnamefont {McCulloch}}, \ and\
  \bibinfo {author} {\bibfnamefont {M.}~\bibnamefont {Roncaglia}},\ }\href@noop
  {} {\bibfield  {journal} {\bibinfo  {journal} {Nat. Commun.}\ }\textbf
  {\bibinfo {volume} {2}},\ \bibinfo {pages} {361} (\bibinfo {year}
  {2011})}\BibitemShut {NoStop}%
\bibitem [{\citenamefont {Guo}\ \emph {et~al.}(2009)\citenamefont {Guo},
  \citenamefont {Hao},\ and\ \citenamefont {Chen}}]{PhysRevA.80.052332}%
  \BibitemOpen
  \bibfield  {author} {\bibinfo {author} {\bibfnamefont {H.}~\bibnamefont
  {Guo}}, \bibinfo {author} {\bibfnamefont {Y.}~\bibnamefont {Hao}}, \ and\
  \bibinfo {author} {\bibfnamefont {S.}~\bibnamefont {Chen}},\ }\href {\doibase
  10.1103/PhysRevA.80.052332} {\bibfield  {journal} {\bibinfo  {journal} {Phys.
  Rev. A}\ }\textbf {\bibinfo {volume} {80}},\ \bibinfo {pages} {052332}
  (\bibinfo {year} {2009})}\BibitemShut {NoStop}%
\bibitem [{\citenamefont {Greschner}\ \emph {et~al.}(2015)\citenamefont
  {Greschner}, \citenamefont {Piraud}, \citenamefont {Heidrich-Meisner},
  \citenamefont {McCulloch}, \citenamefont {Schollw\"ock},\ and\ \citenamefont
  {Vekua}}]{PhysRevLett.115.190402}%
  \BibitemOpen
  \bibfield  {author} {\bibinfo {author} {\bibfnamefont {S.}~\bibnamefont
  {Greschner}}, \bibinfo {author} {\bibfnamefont {M.}~\bibnamefont {Piraud}},
  \bibinfo {author} {\bibfnamefont {F.}~\bibnamefont {Heidrich-Meisner}},
  \bibinfo {author} {\bibfnamefont {I.~P.}\ \bibnamefont {McCulloch}}, \bibinfo
  {author} {\bibfnamefont {U.}~\bibnamefont {Schollw\"ock}}, \ and\ \bibinfo
  {author} {\bibfnamefont {T.}~\bibnamefont {Vekua}},\ }\href {\doibase
  10.1103/PhysRevLett.115.190402} {\bibfield  {journal} {\bibinfo  {journal}
  {Phys. Rev. Lett.}\ }\textbf {\bibinfo {volume} {115}},\ \bibinfo {pages}
  {190402} (\bibinfo {year} {2015})}\BibitemShut {NoStop}%
\bibitem [{\citenamefont {Arcila-Forero}\ \emph {et~al.}(2016)\citenamefont
  {Arcila-Forero}, \citenamefont {Franco},\ and\ \citenamefont
  {Silva-Valencia}}]{PhysRevA.94.013611}%
  \BibitemOpen
  \bibfield  {author} {\bibinfo {author} {\bibfnamefont {J.}~\bibnamefont
  {Arcila-Forero}}, \bibinfo {author} {\bibfnamefont {R.}~\bibnamefont
  {Franco}}, \ and\ \bibinfo {author} {\bibfnamefont {J.}~\bibnamefont
  {Silva-Valencia}},\ }\href {\doibase 10.1103/PhysRevA.94.013611} {\bibfield
  {journal} {\bibinfo  {journal} {Phys. Rev. A}\ }\textbf {\bibinfo {volume}
  {94}},\ \bibinfo {pages} {013611} (\bibinfo {year} {2016})}\BibitemShut
  {NoStop}%
\bibitem [{\citenamefont {Zhang}\ \emph {et~al.}(2017)\citenamefont {Zhang},
  \citenamefont {Greschner}, \citenamefont {Fan}, \citenamefont {Scott},\ and\
  \citenamefont {Zhang}}]{PhysRevA.95.053614}%
  \BibitemOpen
  \bibfield  {author} {\bibinfo {author} {\bibfnamefont {W.}~\bibnamefont
  {Zhang}}, \bibinfo {author} {\bibfnamefont {S.}~\bibnamefont {Greschner}},
  \bibinfo {author} {\bibfnamefont {E.}~\bibnamefont {Fan}}, \bibinfo {author}
  {\bibfnamefont {T.~C.}\ \bibnamefont {Scott}}, \ and\ \bibinfo {author}
  {\bibfnamefont {Y.}~\bibnamefont {Zhang}},\ }\href {\doibase
  10.1103/PhysRevA.95.053614} {\bibfield  {journal} {\bibinfo  {journal} {Phys.
  Rev. A}\ }\textbf {\bibinfo {volume} {95}},\ \bibinfo {pages} {053614}
  (\bibinfo {year} {2017})}\BibitemShut {NoStop}%
\bibitem [{\citenamefont {Greschner}\ and\ \citenamefont
  {Santos}(2015)}]{PhysRevLett.115.053002}%
  \BibitemOpen
  \bibfield  {author} {\bibinfo {author} {\bibfnamefont {S.}~\bibnamefont
  {Greschner}}\ and\ \bibinfo {author} {\bibfnamefont {L.}~\bibnamefont
  {Santos}},\ }\href {\doibase 10.1103/PhysRevLett.115.053002} {\bibfield
  {journal} {\bibinfo  {journal} {Phys. Rev. Lett.}\ }\textbf {\bibinfo
  {volume} {115}},\ \bibinfo {pages} {053002} (\bibinfo {year}
  {2015})}\BibitemShut {NoStop}%
\bibitem [{\citenamefont {Str\"ater}\ \emph {et~al.}(2016)\citenamefont
  {Str\"ater}, \citenamefont {Srivastava},\ and\ \citenamefont
  {Eckardt}}]{PhysRevLett.117.205303}%
  \BibitemOpen
  \bibfield  {author} {\bibinfo {author} {\bibfnamefont {C.}~\bibnamefont
  {Str\"ater}}, \bibinfo {author} {\bibfnamefont {S.~C.~L.}\ \bibnamefont
  {Srivastava}}, \ and\ \bibinfo {author} {\bibfnamefont {A.}~\bibnamefont
  {Eckardt}},\ }\href {\doibase 10.1103/PhysRevLett.117.205303} {\bibfield
  {journal} {\bibinfo  {journal} {Phys. Rev. Lett.}\ }\textbf {\bibinfo
  {volume} {117}},\ \bibinfo {pages} {205303} (\bibinfo {year}
  {2016})}\BibitemShut {NoStop}%
\bibitem [{\citenamefont {Clark}\ \emph {et~al.}(2018)\citenamefont {Clark},
  \citenamefont {Anderson}, \citenamefont {Feng}, \citenamefont {Gaj},
  \citenamefont {Levin},\ and\ \citenamefont {Chin}}]{PhysRevLett.121.030402}%
  \BibitemOpen
  \bibfield  {author} {\bibinfo {author} {\bibfnamefont {L.~W.}\ \bibnamefont
  {Clark}}, \bibinfo {author} {\bibfnamefont {B.~M.}\ \bibnamefont {Anderson}},
  \bibinfo {author} {\bibfnamefont {L.}~\bibnamefont {Feng}}, \bibinfo {author}
  {\bibfnamefont {A.}~\bibnamefont {Gaj}}, \bibinfo {author} {\bibfnamefont
  {K.}~\bibnamefont {Levin}}, \ and\ \bibinfo {author} {\bibfnamefont
  {C.}~\bibnamefont {Chin}},\ }\href {\doibase 10.1103/PhysRevLett.121.030402}
  {\bibfield  {journal} {\bibinfo  {journal} {Phys. Rev. Lett.}\ }\textbf
  {\bibinfo {volume} {121}},\ \bibinfo {pages} {030402} (\bibinfo {year}
  {2018})}\BibitemShut {NoStop}%
\bibitem [{\citenamefont {Yuan}\ \emph {et~al.}(2017)\citenamefont {Yuan},
  \citenamefont {Xiao}, \citenamefont {Xu},\ and\ \citenamefont
  {Fan}}]{PhysRevA.96.043864}%
  \BibitemOpen
  \bibfield  {author} {\bibinfo {author} {\bibfnamefont {L.}~\bibnamefont
  {Yuan}}, \bibinfo {author} {\bibfnamefont {M.}~\bibnamefont {Xiao}}, \bibinfo
  {author} {\bibfnamefont {S.}~\bibnamefont {Xu}}, \ and\ \bibinfo {author}
  {\bibfnamefont {S.}~\bibnamefont {Fan}},\ }\href {\doibase
  10.1103/PhysRevA.96.043864} {\bibfield  {journal} {\bibinfo  {journal} {Phys.
  Rev. A}\ }\textbf {\bibinfo {volume} {96}},\ \bibinfo {pages} {043864}
  (\bibinfo {year} {2017})}\BibitemShut {NoStop}%
\bibitem [{\citenamefont {Lewenstein}\ \emph {et~al.}(2007)\citenamefont
  {Lewenstein}, \citenamefont {Sanpera}, \citenamefont {Ahufinger},
  \citenamefont {Damski}, \citenamefont {Sen},\ and\ \citenamefont
  {Sen}}]{AP56-243}%
  \BibitemOpen
  \bibfield  {author} {\bibinfo {author} {\bibfnamefont {M.}~\bibnamefont
  {Lewenstein}}, \bibinfo {author} {\bibfnamefont {A.}~\bibnamefont {Sanpera}},
  \bibinfo {author} {\bibfnamefont {V.}~\bibnamefont {Ahufinger}}, \bibinfo
  {author} {\bibfnamefont {B.}~\bibnamefont {Damski}}, \bibinfo {author}
  {\bibfnamefont {A.}~\bibnamefont {Sen}}, \ and\ \bibinfo {author}
  {\bibfnamefont {U.}~\bibnamefont {Sen}},\ }\href {\doibase
  10.1080/00018730701223200} {\bibfield  {journal} {\bibinfo  {journal} {Adv.
  Phys.}\ }\textbf {\bibinfo {volume} {56}},\ \bibinfo {pages} {243} (\bibinfo
  {year} {2007})}\BibitemShut {NoStop}%
\bibitem [{\citenamefont {Bloch}\ \emph {et~al.}(2008)\citenamefont {Bloch},
  \citenamefont {Dalibard},\ and\ \citenamefont {Zwerger}}]{RevModPhys.80.885}%
  \BibitemOpen
  \bibfield  {author} {\bibinfo {author} {\bibfnamefont {I.}~\bibnamefont
  {Bloch}}, \bibinfo {author} {\bibfnamefont {J.}~\bibnamefont {Dalibard}}, \
  and\ \bibinfo {author} {\bibfnamefont {W.}~\bibnamefont {Zwerger}},\ }\href
  {\doibase 10.1103/RevModPhys.80.885} {\bibfield  {journal} {\bibinfo
  {journal} {Rev. Mod. Phys.}\ }\textbf {\bibinfo {volume} {80}},\ \bibinfo
  {pages} {885} (\bibinfo {year} {2008})}\BibitemShut {NoStop}%
\bibitem [{\citenamefont {Eisert}\ \emph {et~al.}(2015)\citenamefont {Eisert},
  \citenamefont {Friesdorf},\ and\ \citenamefont {Gogolin}}]{nphys11-124}%
  \BibitemOpen
  \bibfield  {author} {\bibinfo {author} {\bibfnamefont {J.}~\bibnamefont
  {Eisert}}, \bibinfo {author} {\bibfnamefont {M.}~\bibnamefont {Friesdorf}}, \
  and\ \bibinfo {author} {\bibfnamefont {C.}~\bibnamefont {Gogolin}},\
  }\href@noop {} {\bibfield  {journal} {\bibinfo  {journal} {Nat. Phys.}\
  }\textbf {\bibinfo {volume} {11}},\ \bibinfo {pages} {124} (\bibinfo {year}
  {2015})}\BibitemShut {NoStop}%
\bibitem [{\citenamefont {Gogolin}\ and\ \citenamefont
  {Eisert}(2016)}]{Gogolin_2016}%
  \BibitemOpen
  \bibfield  {author} {\bibinfo {author} {\bibfnamefont {C.}~\bibnamefont
  {Gogolin}}\ and\ \bibinfo {author} {\bibfnamefont {J.}~\bibnamefont
  {Eisert}},\ }\href {\doibase 10.1088/0034-4885/79/5/056001} {\bibfield
  {journal} {\bibinfo  {journal} {Rep. Prog. Phys.}\ }\textbf {\bibinfo
  {volume} {79}},\ \bibinfo {pages} {056001} (\bibinfo {year}
  {2016})}\BibitemShut {NoStop}%
\bibitem [{\citenamefont {Ronzheimer}\ \emph {et~al.}(2013)\citenamefont
  {Ronzheimer}, \citenamefont {Schreiber}, \citenamefont {Braun}, \citenamefont
  {Hodgman}, \citenamefont {Langer}, \citenamefont {McCulloch}, \citenamefont
  {Heidrich-Meisner}, \citenamefont {Bloch},\ and\ \citenamefont
  {Schneider}}]{PhysRevLett.110.205301}%
  \BibitemOpen
  \bibfield  {author} {\bibinfo {author} {\bibfnamefont {J.~P.}\ \bibnamefont
  {Ronzheimer}}, \bibinfo {author} {\bibfnamefont {M.}~\bibnamefont
  {Schreiber}}, \bibinfo {author} {\bibfnamefont {S.}~\bibnamefont {Braun}},
  \bibinfo {author} {\bibfnamefont {S.~S.}\ \bibnamefont {Hodgman}}, \bibinfo
  {author} {\bibfnamefont {S.}~\bibnamefont {Langer}}, \bibinfo {author}
  {\bibfnamefont {I.~P.}\ \bibnamefont {McCulloch}}, \bibinfo {author}
  {\bibfnamefont {F.}~\bibnamefont {Heidrich-Meisner}}, \bibinfo {author}
  {\bibfnamefont {I.}~\bibnamefont {Bloch}}, \ and\ \bibinfo {author}
  {\bibfnamefont {U.}~\bibnamefont {Schneider}},\ }\href {\doibase
  10.1103/PhysRevLett.110.205301} {\bibfield  {journal} {\bibinfo  {journal}
  {Phys. Rev. Lett.}\ }\textbf {\bibinfo {volume} {110}},\ \bibinfo {pages}
  {205301} (\bibinfo {year} {2013})}\BibitemShut {NoStop}%
\bibitem [{\citenamefont {Kaufmanm}\ \emph {et~al.}(2016)\citenamefont
  {Kaufmanm}, \citenamefont {Tai}, \citenamefont {Lukin}, \citenamefont
  {Rispoli}, \citenamefont {Schittko}, \citenamefont {Preiss},\ and\
  \citenamefont {Greiner}}]{science353-794}%
  \BibitemOpen
  \bibfield  {author} {\bibinfo {author} {\bibfnamefont {A.~M.}\ \bibnamefont
  {Kaufmanm}}, \bibinfo {author} {\bibfnamefont {M.}~\bibnamefont {Tai}},
  \bibinfo {author} {\bibfnamefont {A.}~\bibnamefont {Lukin}}, \bibinfo
  {author} {\bibfnamefont {M.}~\bibnamefont {Rispoli}}, \bibinfo {author}
  {\bibfnamefont {R.}~\bibnamefont {Schittko}}, \bibinfo {author}
  {\bibfnamefont {P.~M.}\ \bibnamefont {Preiss}}, \ and\ \bibinfo {author}
  {\bibfnamefont {M.}~\bibnamefont {Greiner}},\ }\href@noop {} {\bibfield
  {journal} {\bibinfo  {journal} {Science}\ }\textbf {\bibinfo {volume}
  {353}},\ \bibinfo {pages} {794} (\bibinfo {year} {2016})}\BibitemShut
  {NoStop}%
\bibitem [{\citenamefont {Jurcevic}\ \emph {et~al.}(2017)\citenamefont
  {Jurcevic}, \citenamefont {Shen}, \citenamefont {Hauke}, \citenamefont
  {Maier}, \citenamefont {Brydges}, \citenamefont {Hempel}, \citenamefont
  {Lanyon}, \citenamefont {Heyl}, \citenamefont {Blatt},\ and\ \citenamefont
  {Roos}}]{PhysRevLett.119.080501}%
  \BibitemOpen
  \bibfield  {author} {\bibinfo {author} {\bibfnamefont {P.}~\bibnamefont
  {Jurcevic}}, \bibinfo {author} {\bibfnamefont {H.}~\bibnamefont {Shen}},
  \bibinfo {author} {\bibfnamefont {P.}~\bibnamefont {Hauke}}, \bibinfo
  {author} {\bibfnamefont {C.}~\bibnamefont {Maier}}, \bibinfo {author}
  {\bibfnamefont {T.}~\bibnamefont {Brydges}}, \bibinfo {author} {\bibfnamefont
  {C.}~\bibnamefont {Hempel}}, \bibinfo {author} {\bibfnamefont {B.~P.}\
  \bibnamefont {Lanyon}}, \bibinfo {author} {\bibfnamefont {M.}~\bibnamefont
  {Heyl}}, \bibinfo {author} {\bibfnamefont {R.}~\bibnamefont {Blatt}}, \ and\
  \bibinfo {author} {\bibfnamefont {C.~F.}\ \bibnamefont {Roos}},\ }\href
  {\doibase 10.1103/PhysRevLett.119.080501} {\bibfield  {journal} {\bibinfo
  {journal} {Phys. Rev. Lett.}\ }\textbf {\bibinfo {volume} {119}},\ \bibinfo
  {pages} {080501} (\bibinfo {year} {2017})}\BibitemShut {NoStop}%
\bibitem [{\citenamefont {Wright}\ \emph {et~al.}(2014)\citenamefont {Wright},
  \citenamefont {Rigol}, \citenamefont {Davis},\ and\ \citenamefont
  {Kheruntsyan}}]{PhysRevLett.113.050601}%
  \BibitemOpen
  \bibfield  {author} {\bibinfo {author} {\bibfnamefont {T.~M.}\ \bibnamefont
  {Wright}}, \bibinfo {author} {\bibfnamefont {M.}~\bibnamefont {Rigol}},
  \bibinfo {author} {\bibfnamefont {M.~J.}\ \bibnamefont {Davis}}, \ and\
  \bibinfo {author} {\bibfnamefont {K.~V.}\ \bibnamefont {Kheruntsyan}},\
  }\href {\doibase 10.1103/PhysRevLett.113.050601} {\bibfield  {journal}
  {\bibinfo  {journal} {Phys. Rev. Lett.}\ }\textbf {\bibinfo {volume} {113}},\
  \bibinfo {pages} {050601} (\bibinfo {year} {2014})}\BibitemShut {NoStop}%
\bibitem [{\citenamefont {Lange}\ \emph
  {et~al.}(2017{\natexlab{b}})\citenamefont {Lange}, \citenamefont {Ejima},\
  and\ \citenamefont {Fehske}}]{PhysRevA.95.063621}%
  \BibitemOpen
  \bibfield  {author} {\bibinfo {author} {\bibfnamefont {F.}~\bibnamefont
  {Lange}}, \bibinfo {author} {\bibfnamefont {S.}~\bibnamefont {Ejima}}, \ and\
  \bibinfo {author} {\bibfnamefont {H.}~\bibnamefont {Fehske}},\ }\href
  {\doibase 10.1103/PhysRevA.95.063621} {\bibfield  {journal} {\bibinfo
  {journal} {Phys. Rev. A}\ }\textbf {\bibinfo {volume} {95}},\ \bibinfo
  {pages} {063621} (\bibinfo {year} {2017}{\natexlab{b}})}\BibitemShut
  {NoStop}%
\bibitem [{\citenamefont {Settino}\ \emph {et~al.}(2021)\citenamefont
  {Settino}, \citenamefont {Lo~Gullo}, \citenamefont {Plastina},\ and\
  \citenamefont {Minguzzi}}]{PhysRevLett.126.065301}%
  \BibitemOpen
  \bibfield  {author} {\bibinfo {author} {\bibfnamefont {J.}~\bibnamefont
  {Settino}}, \bibinfo {author} {\bibfnamefont {N.}~\bibnamefont {Lo~Gullo}},
  \bibinfo {author} {\bibfnamefont {F.}~\bibnamefont {Plastina}}, \ and\
  \bibinfo {author} {\bibfnamefont {A.}~\bibnamefont {Minguzzi}},\ }\href
  {\doibase 10.1103/PhysRevLett.126.065301} {\bibfield  {journal} {\bibinfo
  {journal} {Phys. Rev. Lett.}\ }\textbf {\bibinfo {volume} {126}},\ \bibinfo
  {pages} {065301} (\bibinfo {year} {2021})}\BibitemShut {NoStop}%
\bibitem [{\citenamefont {Liu}\ \emph {et~al.}(2018)\citenamefont {Liu},
  \citenamefont {Garrison}, \citenamefont {Deng}, \citenamefont {Gong},\ and\
  \citenamefont {Gorshkov}}]{PhysRevLett.121.250404}%
  \BibitemOpen
  \bibfield  {author} {\bibinfo {author} {\bibfnamefont {F.}~\bibnamefont
  {Liu}}, \bibinfo {author} {\bibfnamefont {J.~R.}\ \bibnamefont {Garrison}},
  \bibinfo {author} {\bibfnamefont {D.-L.}\ \bibnamefont {Deng}}, \bibinfo
  {author} {\bibfnamefont {Z.-X.}\ \bibnamefont {Gong}}, \ and\ \bibinfo
  {author} {\bibfnamefont {A.~V.}\ \bibnamefont {Gorshkov}},\ }\href {\doibase
  10.1103/PhysRevLett.121.250404} {\bibfield  {journal} {\bibinfo  {journal}
  {Phys. Rev. Lett.}\ }\textbf {\bibinfo {volume} {121}},\ \bibinfo {pages}
  {250404} (\bibinfo {year} {2018})}\BibitemShut {NoStop}%
\bibitem [{\citenamefont {Damascelli}(2004)}]{Damascelli_2004}%
  \BibitemOpen
  \bibfield  {author} {\bibinfo {author} {\bibfnamefont {A.}~\bibnamefont
  {Damascelli}},\ }\href {\doibase 10.1238/physica.topical.109a00061}
  {\bibfield  {journal} {\bibinfo  {journal} {Phys. Scr.}\ }\textbf {\bibinfo
  {volume} {T109}},\ \bibinfo {pages} {61} (\bibinfo {year}
  {2004})}\BibitemShut {NoStop}%
\bibitem [{\citenamefont {Stewart}\ \emph {et~al.}(2008)\citenamefont
  {Stewart}, \citenamefont {Gaebler},\ and\ \citenamefont
  {Jin}}]{nature454-744}%
  \BibitemOpen
  \bibfield  {author} {\bibinfo {author} {\bibfnamefont {J.}~\bibnamefont
  {Stewart}}, \bibinfo {author} {\bibfnamefont {J.}~\bibnamefont {Gaebler}}, \
  and\ \bibinfo {author} {\bibfnamefont {D.}~\bibnamefont {Jin}},\ }\href@noop
  {} {\bibfield  {journal} {\bibinfo  {journal} {Nature}\ }\textbf {\bibinfo
  {volume} {454}},\ \bibinfo {pages} {744} (\bibinfo {year}
  {2008})}\BibitemShut {NoStop}%
\bibitem [{\citenamefont {Volchkov}\ \emph {et~al.}(2018)\citenamefont
  {Volchkov}, \citenamefont {Pasek}, \citenamefont {Denechaud}, \citenamefont
  {Mukhtar}, \citenamefont {Aspect}, \citenamefont {Delande},\ and\
  \citenamefont {Josse}}]{PhysRevLett.120.060404}%
  \BibitemOpen
  \bibfield  {author} {\bibinfo {author} {\bibfnamefont {V.~V.}\ \bibnamefont
  {Volchkov}}, \bibinfo {author} {\bibfnamefont {M.}~\bibnamefont {Pasek}},
  \bibinfo {author} {\bibfnamefont {V.}~\bibnamefont {Denechaud}}, \bibinfo
  {author} {\bibfnamefont {M.}~\bibnamefont {Mukhtar}}, \bibinfo {author}
  {\bibfnamefont {A.}~\bibnamefont {Aspect}}, \bibinfo {author} {\bibfnamefont
  {D.}~\bibnamefont {Delande}}, \ and\ \bibinfo {author} {\bibfnamefont
  {V.}~\bibnamefont {Josse}},\ }\href {\doibase 10.1103/PhysRevLett.120.060404}
  {\bibfield  {journal} {\bibinfo  {journal} {Phys. Rev. Lett.}\ }\textbf
  {\bibinfo {volume} {120}},\ \bibinfo {pages} {060404} (\bibinfo {year}
  {2018})}\BibitemShut {NoStop}%
\bibitem [{\citenamefont {Bohrdt}\ \emph {et~al.}(2018)\citenamefont {Bohrdt},
  \citenamefont {Greif}, \citenamefont {Demler}, \citenamefont {Knap},\ and\
  \citenamefont {Grusdt}}]{PhysRevB.97.125117}%
  \BibitemOpen
  \bibfield  {author} {\bibinfo {author} {\bibfnamefont {A.}~\bibnamefont
  {Bohrdt}}, \bibinfo {author} {\bibfnamefont {D.}~\bibnamefont {Greif}},
  \bibinfo {author} {\bibfnamefont {E.}~\bibnamefont {Demler}}, \bibinfo
  {author} {\bibfnamefont {M.}~\bibnamefont {Knap}}, \ and\ \bibinfo {author}
  {\bibfnamefont {F.}~\bibnamefont {Grusdt}},\ }\href {\doibase
  10.1103/PhysRevB.97.125117} {\bibfield  {journal} {\bibinfo  {journal} {Phys.
  Rev. B}\ }\textbf {\bibinfo {volume} {97}},\ \bibinfo {pages} {125117}
  (\bibinfo {year} {2018})}\BibitemShut {NoStop}%
\bibitem [{\citenamefont {Roberts}\ and\ \citenamefont
  {Stanford}(2015)}]{PhysRevLett.115.131603}%
  \BibitemOpen
  \bibfield  {author} {\bibinfo {author} {\bibfnamefont {D.~A.}\ \bibnamefont
  {Roberts}}\ and\ \bibinfo {author} {\bibfnamefont {D.}~\bibnamefont
  {Stanford}},\ }\href {\doibase 10.1103/PhysRevLett.115.131603} {\bibfield
  {journal} {\bibinfo  {journal} {Phys. Rev. Lett.}\ }\textbf {\bibinfo
  {volume} {115}},\ \bibinfo {pages} {131603} (\bibinfo {year}
  {2015})}\BibitemShut {NoStop}%
\bibitem [{\citenamefont {Polchinski}\ and\ \citenamefont
  {Rosenhaus}(2016)}]{JHEP2016-1}%
  \BibitemOpen
  \bibfield  {author} {\bibinfo {author} {\bibfnamefont {J.}~\bibnamefont
  {Polchinski}}\ and\ \bibinfo {author} {\bibfnamefont {V.}~\bibnamefont
  {Rosenhaus}},\ }\href {\doibase 10.1007/JHEP04(2016)001} {\bibfield
  {journal} {\bibinfo  {journal} {J. High Energ. Phys.}\ }\textbf {\bibinfo
  {volume} {2016}},\ \bibinfo {pages} {1} (\bibinfo {year} {2016})}\BibitemShut
  {NoStop}%
\bibitem [{\citenamefont {Maldacena}\ \emph {et~al.}(2016)\citenamefont
  {Maldacena}, \citenamefont {Shenker},\ and\ \citenamefont
  {Stanford}}]{JHEP2016-106}%
  \BibitemOpen
  \bibfield  {author} {\bibinfo {author} {\bibfnamefont {J.}~\bibnamefont
  {Maldacena}}, \bibinfo {author} {\bibfnamefont {S.}~\bibnamefont {Shenker}},
  \ and\ \bibinfo {author} {\bibfnamefont {D.}~\bibnamefont {Stanford}},\
  }\href {\doibase 10.1007/JHEP08(2016)106} {\bibfield  {journal} {\bibinfo
  {journal} {J. High Energ. Phys.}\ }\textbf {\bibinfo {volume} {2016}},\
  \bibinfo {pages} {106} (\bibinfo {year} {2016})}\BibitemShut {NoStop}%
\bibitem [{\citenamefont {Gu}\ and\ \citenamefont {Qi}(2016)}]{JHEP2016-129}%
  \BibitemOpen
  \bibfield  {author} {\bibinfo {author} {\bibfnamefont {Y.}~\bibnamefont
  {Gu}}\ and\ \bibinfo {author} {\bibfnamefont {X.~L.}\ \bibnamefont {Qi}},\
  }\href {\doibase 10.1007/JHEP08(2016)129} {\bibfield  {journal} {\bibinfo
  {journal} {J. High Energ. Phys.}\ }\textbf {\bibinfo {volume} {2016}},\
  \bibinfo {pages} {129} (\bibinfo {year} {2016})}\BibitemShut {NoStop}%
\bibitem [{\citenamefont {Mezei}\ and\ \citenamefont
  {Stanford}(2017)}]{JHEP2017-65}%
  \BibitemOpen
  \bibfield  {author} {\bibinfo {author} {\bibfnamefont {M.}~\bibnamefont
  {Mezei}}\ and\ \bibinfo {author} {\bibfnamefont {D.}~\bibnamefont
  {Stanford}},\ }\href {\doibase 10.1007/JHEP05(2017)065} {\bibfield  {journal}
  {\bibinfo  {journal} {J. High Energ. Phys.}\ }\textbf {\bibinfo {volume}
  {2017}},\ \bibinfo {pages} {65} (\bibinfo {year} {2017})}\BibitemShut
  {NoStop}%
\bibitem [{\citenamefont {Styliaris}\ \emph {et~al.}(2021)\citenamefont
  {Styliaris}, \citenamefont {Anand},\ and\ \citenamefont
  {Zanardi}}]{PhysRevLett.126.030601}%
  \BibitemOpen
  \bibfield  {author} {\bibinfo {author} {\bibfnamefont {G.}~\bibnamefont
  {Styliaris}}, \bibinfo {author} {\bibfnamefont {N.}~\bibnamefont {Anand}}, \
  and\ \bibinfo {author} {\bibfnamefont {P.}~\bibnamefont {Zanardi}},\ }\href
  {\doibase 10.1103/PhysRevLett.126.030601} {\bibfield  {journal} {\bibinfo
  {journal} {Phys. Rev. Lett.}\ }\textbf {\bibinfo {volume} {126}},\ \bibinfo
  {pages} {030601} (\bibinfo {year} {2021})}\BibitemShut {NoStop}%
\bibitem [{\citenamefont {Zanardi}\ and\ \citenamefont
  {Anand}(2021)}]{PhysRevA.103.062214}%
  \BibitemOpen
  \bibfield  {author} {\bibinfo {author} {\bibfnamefont {P.}~\bibnamefont
  {Zanardi}}\ and\ \bibinfo {author} {\bibfnamefont {N.}~\bibnamefont
  {Anand}},\ }\href {\doibase 10.1103/PhysRevA.103.062214} {\bibfield
  {journal} {\bibinfo  {journal} {Phys. Rev. A}\ }\textbf {\bibinfo {volume}
  {103}},\ \bibinfo {pages} {062214} (\bibinfo {year} {2021})}\BibitemShut
  {NoStop}%
\bibitem [{\citenamefont {Da\ifmmode~\breve{g}\else \u{g}\fi{}}\ \emph
  {et~al.}(2019)\citenamefont {Da\ifmmode~\breve{g}\else \u{g}\fi{}},
  \citenamefont {Sun},\ and\ \citenamefont {Duan}}]{PhysRevLett.123.140602}%
  \BibitemOpen
  \bibfield  {author} {\bibinfo {author} {\bibfnamefont {C.~B.}\ \bibnamefont
  {Da\ifmmode~\breve{g}\else \u{g}\fi{}}}, \bibinfo {author} {\bibfnamefont
  {K.}~\bibnamefont {Sun}}, \ and\ \bibinfo {author} {\bibfnamefont {L.-M.}\
  \bibnamefont {Duan}},\ }\href {\doibase 10.1103/PhysRevLett.123.140602}
  {\bibfield  {journal} {\bibinfo  {journal} {Phys. Rev. Lett.}\ }\textbf
  {\bibinfo {volume} {123}},\ \bibinfo {pages} {140602} (\bibinfo {year}
  {2019})}\BibitemShut {NoStop}%
\bibitem [{\citenamefont {Huang}\ \emph {et~al.}(2017)\citenamefont {Huang},
  \citenamefont {Zhang},\ and\ \citenamefont {Chen}}]{AdP529-1600318}%
  \BibitemOpen
  \bibfield  {author} {\bibinfo {author} {\bibfnamefont {Y.}~\bibnamefont
  {Huang}}, \bibinfo {author} {\bibfnamefont {Y.-L.}\ \bibnamefont {Zhang}}, \
  and\ \bibinfo {author} {\bibfnamefont {X.}~\bibnamefont {Chen}},\ }\href
  {\doibase https://doi.org/10.1002/andp.201600318} {\bibfield  {journal}
  {\bibinfo  {journal} {Annalen der Physik}\ }\textbf {\bibinfo {volume}
  {529}},\ \bibinfo {pages} {1600318} (\bibinfo {year} {2017})}\BibitemShut
  {NoStop}%
\bibitem [{\citenamefont {Chen}\ \emph {et~al.}(2017)\citenamefont {Chen},
  \citenamefont {Zhou}, \citenamefont {Huse},\ and\ \citenamefont
  {Fradkin}}]{AdP529-1600332}%
  \BibitemOpen
  \bibfield  {author} {\bibinfo {author} {\bibfnamefont {X.}~\bibnamefont
  {Chen}}, \bibinfo {author} {\bibfnamefont {T.}~\bibnamefont {Zhou}}, \bibinfo
  {author} {\bibfnamefont {D.~A.}\ \bibnamefont {Huse}}, \ and\ \bibinfo
  {author} {\bibfnamefont {E.}~\bibnamefont {Fradkin}},\ }\href {\doibase
  https://doi.org/10.1002/andp.201600332} {\bibfield  {journal} {\bibinfo
  {journal} {Annalen der Physik}\ }\textbf {\bibinfo {volume} {529}},\ \bibinfo
  {pages} {1600332} (\bibinfo {year} {2017})}\BibitemShut {NoStop}%
\bibitem [{\citenamefont {Fan}\ \emph {et~al.}(2017)\citenamefont {Fan},
  \citenamefont {Zhang}, \citenamefont {Shen},\ and\ \citenamefont
  {Zhai}}]{FAN2017707}%
  \BibitemOpen
  \bibfield  {author} {\bibinfo {author} {\bibfnamefont {R.}~\bibnamefont
  {Fan}}, \bibinfo {author} {\bibfnamefont {P.}~\bibnamefont {Zhang}}, \bibinfo
  {author} {\bibfnamefont {H.}~\bibnamefont {Shen}}, \ and\ \bibinfo {author}
  {\bibfnamefont {H.}~\bibnamefont {Zhai}},\ }\href {\doibase
  https://doi.org/10.1016/j.scib.2017.04.011} {\bibfield  {journal} {\bibinfo
  {journal} {Science Bulletin}\ }\textbf {\bibinfo {volume} {62}},\ \bibinfo
  {pages} {707} (\bibinfo {year} {2017})}\BibitemShut {NoStop}%
\bibitem [{\citenamefont {He}\ and\ \citenamefont
  {Lu}(2017)}]{PhysRevB.95.054201}%
  \BibitemOpen
  \bibfield  {author} {\bibinfo {author} {\bibfnamefont {R.-Q.}\ \bibnamefont
  {He}}\ and\ \bibinfo {author} {\bibfnamefont {Z.-Y.}\ \bibnamefont {Lu}},\
  }\href {\doibase 10.1103/PhysRevB.95.054201} {\bibfield  {journal} {\bibinfo
  {journal} {Phys. Rev. B}\ }\textbf {\bibinfo {volume} {95}},\ \bibinfo
  {pages} {054201} (\bibinfo {year} {2017})}\BibitemShut {NoStop}%
\bibitem [{\citenamefont {Swingle}\ and\ \citenamefont
  {Chowdhury}(2017)}]{PhysRevB.95.060201}%
  \BibitemOpen
  \bibfield  {author} {\bibinfo {author} {\bibfnamefont {B.}~\bibnamefont
  {Swingle}}\ and\ \bibinfo {author} {\bibfnamefont {D.}~\bibnamefont
  {Chowdhury}},\ }\href {\doibase 10.1103/PhysRevB.95.060201} {\bibfield
  {journal} {\bibinfo  {journal} {Phys. Rev. B}\ }\textbf {\bibinfo {volume}
  {95}},\ \bibinfo {pages} {060201} (\bibinfo {year} {2017})}\BibitemShut
  {NoStop}%
\bibitem [{\citenamefont {Zvonarev}\ \emph {et~al.}(2009)\citenamefont
  {Zvonarev}, \citenamefont {Cheianov},\ and\ \citenamefont
  {Giamarchi}}]{Zvonarev_2009}%
  \BibitemOpen
  \bibfield  {author} {\bibinfo {author} {\bibfnamefont {M.~B.}\ \bibnamefont
  {Zvonarev}}, \bibinfo {author} {\bibfnamefont {V.~V.}\ \bibnamefont
  {Cheianov}}, \ and\ \bibinfo {author} {\bibfnamefont {T.}~\bibnamefont
  {Giamarchi}},\ }\href {\doibase 10.1088/1742-5468/2009/07/p07035} {\bibfield
  {journal} {\bibinfo  {journal} {J. Stat. Mech. Theory Exp.}\ }\textbf
  {\bibinfo {volume} {2009}},\ \bibinfo {pages} {P07035} (\bibinfo {year}
  {2009})}\BibitemShut {NoStop}%
\bibitem [{\citenamefont {P{\^{a}}{\c{t}}u}\ \emph {et~al.}(2008)\citenamefont
  {P{\^{a}}{\c{t}}u}, \citenamefont {Korepin},\ and\ \citenamefont
  {Averin}}]{Pu_2008b}%
  \BibitemOpen
  \bibfield  {author} {\bibinfo {author} {\bibfnamefont {O.~I.}\ \bibnamefont
  {P{\^{a}}{\c{t}}u}}, \bibinfo {author} {\bibfnamefont {V.~E.}\ \bibnamefont
  {Korepin}}, \ and\ \bibinfo {author} {\bibfnamefont {D.~V.}\ \bibnamefont
  {Averin}},\ }\href {\doibase 10.1088/1751-8113/41/25/255205} {\bibfield
  {journal} {\bibinfo  {journal} {J. Phys. A}\ }\textbf {\bibinfo {volume}
  {41}},\ \bibinfo {pages} {255205} (\bibinfo {year} {2008})}\BibitemShut
  {NoStop}%
\bibitem [{\citenamefont {Zhuravlev}\ \emph {et~al.}(2021)\citenamefont
  {Zhuravlev}, \citenamefont {Naichuk}, \citenamefont {Iorgov},\ and\
  \citenamefont {Gamayun}}]{2110.06860}%
  \BibitemOpen
  \bibfield  {author} {\bibinfo {author} {\bibfnamefont {Y.}~\bibnamefont
  {Zhuravlev}}, \bibinfo {author} {\bibfnamefont {E.}~\bibnamefont {Naichuk}},
  \bibinfo {author} {\bibfnamefont {N.}~\bibnamefont {Iorgov}}, \ and\ \bibinfo
  {author} {\bibfnamefont {O.}~\bibnamefont {Gamayun}},\ }\href@noop {}
  {\enquote {\bibinfo {title} {Large time and long distance asymptotics of the
  thermal correlators of the impenetrable anyonic lattice gas},}\ }\bibinfo
  {howpublished} {arXiv: 2110.06860} (\bibinfo {year} {2021})\BibitemShut
  {NoStop}%
\bibitem [{\citenamefont {Lieb}(1963)}]{PhysRev.130.1616}%
  \BibitemOpen
  \bibfield  {author} {\bibinfo {author} {\bibfnamefont {E.~H.}\ \bibnamefont
  {Lieb}},\ }\href {\doibase 10.1103/PhysRev.130.1616} {\bibfield  {journal}
  {\bibinfo  {journal} {Phys. Rev.}\ }\textbf {\bibinfo {volume} {130}},\
  \bibinfo {pages} {1616} (\bibinfo {year} {1963})}\BibitemShut {NoStop}%
\bibitem [{\citenamefont {Gamayun}\ \emph {et~al.}(2020)\citenamefont
  {Gamayun}, \citenamefont {Lychkovskiy},\ and\ \citenamefont
  {Zvonarev}}]{scipost8-053}%
  \BibitemOpen
  \bibfield  {author} {\bibinfo {author} {\bibfnamefont {O.}~\bibnamefont
  {Gamayun}}, \bibinfo {author} {\bibfnamefont {O.}~\bibnamefont
  {Lychkovskiy}}, \ and\ \bibinfo {author} {\bibfnamefont {M.~B.}\ \bibnamefont
  {Zvonarev}},\ }\href {\doibase 10.21468/SciPostPhys.8.4.053} {\bibfield
  {journal} {\bibinfo  {journal} {SciPost Phys.}\ }\textbf {\bibinfo {volume}
  {8}},\ \bibinfo {pages} {053} (\bibinfo {year} {2020})}\BibitemShut {NoStop}%
\bibitem [{\citenamefont {Olshanii}\ and\ \citenamefont
  {Dunjko}(2003)}]{PhysRevLett.91.090401}%
  \BibitemOpen
  \bibfield  {author} {\bibinfo {author} {\bibfnamefont {M.}~\bibnamefont
  {Olshanii}}\ and\ \bibinfo {author} {\bibfnamefont {V.}~\bibnamefont
  {Dunjko}},\ }\href {\doibase 10.1103/PhysRevLett.91.090401} {\bibfield
  {journal} {\bibinfo  {journal} {Phys. Rev. Lett.}\ }\textbf {\bibinfo
  {volume} {91}},\ \bibinfo {pages} {090401} (\bibinfo {year}
  {2003})}\BibitemShut {NoStop}%
\bibitem [{\citenamefont {Kollath}\ \emph {et~al.}(2007)\citenamefont
  {Kollath}, \citenamefont {K\"ohl},\ and\ \citenamefont
  {Giamarchi}}]{PhysRevA.76.063602}%
  \BibitemOpen
  \bibfield  {author} {\bibinfo {author} {\bibfnamefont {C.}~\bibnamefont
  {Kollath}}, \bibinfo {author} {\bibfnamefont {M.}~\bibnamefont {K\"ohl}}, \
  and\ \bibinfo {author} {\bibfnamefont {T.}~\bibnamefont {Giamarchi}},\ }\href
  {\doibase 10.1103/PhysRevA.76.063602} {\bibfield  {journal} {\bibinfo
  {journal} {Phys. Rev. A}\ }\textbf {\bibinfo {volume} {76}},\ \bibinfo
  {pages} {063602} (\bibinfo {year} {2007})}\BibitemShut {NoStop}%
\bibitem [{\citenamefont {Papi\ifmmode~\acute{c}\else \'{c}\fi{}}\ \emph
  {et~al.}(2018)\citenamefont {Papi\ifmmode~\acute{c}\else \'{c}\fi{}},
  \citenamefont {Mong}, \citenamefont {Yazdani},\ and\ \citenamefont
  {Zaletel}}]{PhysRevX.8.011037}%
  \BibitemOpen
  \bibfield  {author} {\bibinfo {author} {\bibfnamefont {Z.}~\bibnamefont
  {Papi\ifmmode~\acute{c}\else \'{c}\fi{}}}, \bibinfo {author} {\bibfnamefont
  {R.~S.~K.}\ \bibnamefont {Mong}}, \bibinfo {author} {\bibfnamefont
  {A.}~\bibnamefont {Yazdani}}, \ and\ \bibinfo {author} {\bibfnamefont
  {M.~P.}\ \bibnamefont {Zaletel}},\ }\href {\doibase
  10.1103/PhysRevX.8.011037} {\bibfield  {journal} {\bibinfo  {journal} {Phys.
  Rev. X}\ }\textbf {\bibinfo {volume} {8}},\ \bibinfo {pages} {011037}
  (\bibinfo {year} {2018})}\BibitemShut {NoStop}%
\bibitem [{\citenamefont {Imambekov}\ and\ \citenamefont
  {Glazman}(2009{\natexlab{a}})}]{science323-228}%
  \BibitemOpen
  \bibfield  {author} {\bibinfo {author} {\bibfnamefont {A.}~\bibnamefont
  {Imambekov}}\ and\ \bibinfo {author} {\bibfnamefont {L.~I.}\ \bibnamefont
  {Glazman}},\ }\href {\doibase 10.1126/science.1165403} {\bibfield  {journal}
  {\bibinfo  {journal} {Science}\ }\textbf {\bibinfo {volume} {323}},\ \bibinfo
  {pages} {228} (\bibinfo {year} {2009}{\natexlab{a}})}\BibitemShut {NoStop}%
\bibitem [{\citenamefont {Imambekov}\ \emph {et~al.}(2012)\citenamefont
  {Imambekov}, \citenamefont {Schmidt},\ and\ \citenamefont
  {Glazman}}]{RevModPhys.84.1253}%
  \BibitemOpen
  \bibfield  {author} {\bibinfo {author} {\bibfnamefont {A.}~\bibnamefont
  {Imambekov}}, \bibinfo {author} {\bibfnamefont {T.~L.}\ \bibnamefont
  {Schmidt}}, \ and\ \bibinfo {author} {\bibfnamefont {L.~I.}\ \bibnamefont
  {Glazman}},\ }\href {\doibase 10.1103/RevModPhys.84.1253} {\bibfield
  {journal} {\bibinfo  {journal} {Rev. Mod. Phys.}\ }\textbf {\bibinfo {volume}
  {84}},\ \bibinfo {pages} {1253} (\bibinfo {year} {2012})}\BibitemShut
  {NoStop}%
\bibitem [{\citenamefont {Imambekov}\ and\ \citenamefont
  {Glazman}(2008)}]{PhysRevLett.100.206805}%
  \BibitemOpen
  \bibfield  {author} {\bibinfo {author} {\bibfnamefont {A.}~\bibnamefont
  {Imambekov}}\ and\ \bibinfo {author} {\bibfnamefont {L.~I.}\ \bibnamefont
  {Glazman}},\ }\href {\doibase 10.1103/PhysRevLett.100.206805} {\bibfield
  {journal} {\bibinfo  {journal} {Phys. Rev. Lett.}\ }\textbf {\bibinfo
  {volume} {100}},\ \bibinfo {pages} {206805} (\bibinfo {year}
  {2008})}\BibitemShut {NoStop}%
\bibitem [{\citenamefont {Imambekov}\ and\ \citenamefont
  {Glazman}(2009{\natexlab{b}})}]{PhysRevLett.102.126405}%
  \BibitemOpen
  \bibfield  {author} {\bibinfo {author} {\bibfnamefont {A.}~\bibnamefont
  {Imambekov}}\ and\ \bibinfo {author} {\bibfnamefont {L.~I.}\ \bibnamefont
  {Glazman}},\ }\href {\doibase 10.1103/PhysRevLett.102.126405} {\bibfield
  {journal} {\bibinfo  {journal} {Phys. Rev. Lett.}\ }\textbf {\bibinfo
  {volume} {102}},\ \bibinfo {pages} {126405} (\bibinfo {year}
  {2009}{\natexlab{b}})}\BibitemShut {NoStop}%
\bibitem [{\citenamefont {Campbell}\ and\ \citenamefont
  {Gangardt}(2017)}]{scipost3-015}%
  \BibitemOpen
  \bibfield  {author} {\bibinfo {author} {\bibfnamefont {A.}~\bibnamefont
  {Campbell}}\ and\ \bibinfo {author} {\bibfnamefont {D.}~\bibnamefont
  {Gangardt}},\ }\href {\doibase 10.21468/SciPostPhys.3.2.015} {\bibfield
  {journal} {\bibinfo  {journal} {SciPost Phys.}\ }\textbf {\bibinfo {volume}
  {3}},\ \bibinfo {pages} {015} (\bibinfo {year} {2017})}\BibitemShut {NoStop}%
\bibitem [{\citenamefont {Cheneau}\ \emph {et~al.}(2012)\citenamefont
  {Cheneau}, \citenamefont {Barmettler}, \citenamefont {Poletti}, \citenamefont
  {Endres}, \citenamefont {Schaub}, \citenamefont {Fukuhara}, \citenamefont
  {Gross}, \citenamefont {Bloch}, \citenamefont {Kollath},\ and\ \citenamefont
  {Kuhr}}]{nature481-484}%
  \BibitemOpen
  \bibfield  {author} {\bibinfo {author} {\bibfnamefont {M.}~\bibnamefont
  {Cheneau}}, \bibinfo {author} {\bibfnamefont {P.}~\bibnamefont {Barmettler}},
  \bibinfo {author} {\bibfnamefont {D.}~\bibnamefont {Poletti}}, \bibinfo
  {author} {\bibfnamefont {M.}~\bibnamefont {Endres}}, \bibinfo {author}
  {\bibfnamefont {P.}~\bibnamefont {Schaub}}, \bibinfo {author} {\bibfnamefont
  {T.}~\bibnamefont {Fukuhara}}, \bibinfo {author} {\bibfnamefont
  {C.}~\bibnamefont {Gross}}, \bibinfo {author} {\bibfnamefont
  {I.}~\bibnamefont {Bloch}}, \bibinfo {author} {\bibfnamefont
  {C.}~\bibnamefont {Kollath}}, \ and\ \bibinfo {author} {\bibfnamefont
  {S.}~\bibnamefont {Kuhr}},\ }\href {\doibase 10.1038/nature10748} {\bibfield
  {journal} {\bibinfo  {journal} {Nature}\ }\textbf {\bibinfo {volume} {481}},\
  \bibinfo {pages} {484} (\bibinfo {year} {2012})}\BibitemShut {NoStop}%
\bibitem [{\citenamefont {Luitz}\ and\ \citenamefont
  {Bar~Lev}(2017)}]{PhysRevB.96.020406}%
  \BibitemOpen
  \bibfield  {author} {\bibinfo {author} {\bibfnamefont {D.~J.}\ \bibnamefont
  {Luitz}}\ and\ \bibinfo {author} {\bibfnamefont {Y.}~\bibnamefont
  {Bar~Lev}},\ }\href {\doibase 10.1103/PhysRevB.96.020406} {\bibfield
  {journal} {\bibinfo  {journal} {Phys. Rev. B}\ }\textbf {\bibinfo {volume}
  {96}},\ \bibinfo {pages} {020406} (\bibinfo {year} {2017})}\BibitemShut
  {NoStop}%
\bibitem [{\citenamefont {Shen}\ \emph {et~al.}(2017)\citenamefont {Shen},
  \citenamefont {Zhang}, \citenamefont {Fan},\ and\ \citenamefont
  {Zhai}}]{PhysRevB.96.054503}%
  \BibitemOpen
  \bibfield  {author} {\bibinfo {author} {\bibfnamefont {H.}~\bibnamefont
  {Shen}}, \bibinfo {author} {\bibfnamefont {P.}~\bibnamefont {Zhang}},
  \bibinfo {author} {\bibfnamefont {R.}~\bibnamefont {Fan}}, \ and\ \bibinfo
  {author} {\bibfnamefont {H.}~\bibnamefont {Zhai}},\ }\href {\doibase
  10.1103/PhysRevB.96.054503} {\bibfield  {journal} {\bibinfo  {journal} {Phys.
  Rev. B}\ }\textbf {\bibinfo {volume} {96}},\ \bibinfo {pages} {054503}
  (\bibinfo {year} {2017})}\BibitemShut {NoStop}%
\bibitem [{\citenamefont {Bohrdt}\ \emph {et~al.}(2017)\citenamefont {Bohrdt},
  \citenamefont {Mendl}, \citenamefont {Endres},\ and\ \citenamefont
  {Knap}}]{Bohrdt_2017}%
  \BibitemOpen
  \bibfield  {author} {\bibinfo {author} {\bibfnamefont {A.}~\bibnamefont
  {Bohrdt}}, \bibinfo {author} {\bibfnamefont {C.~B.}\ \bibnamefont {Mendl}},
  \bibinfo {author} {\bibfnamefont {M.}~\bibnamefont {Endres}}, \ and\ \bibinfo
  {author} {\bibfnamefont {M.}~\bibnamefont {Knap}},\ }\href {\doibase
  10.1088/1367-2630/aa719b} {\bibfield  {journal} {\bibinfo  {journal} {New J.
  Phys.}\ }\textbf {\bibinfo {volume} {19}},\ \bibinfo {pages} {063001}
  (\bibinfo {year} {2017})}\BibitemShut {NoStop}%
\bibitem [{\citenamefont {Zhang}\ and\ \citenamefont
  {Khemani}(2020)}]{scipost9-024}%
  \BibitemOpen
  \bibfield  {author} {\bibinfo {author} {\bibfnamefont {Y.-L.}\ \bibnamefont
  {Zhang}}\ and\ \bibinfo {author} {\bibfnamefont {V.}~\bibnamefont
  {Khemani}},\ }\href {\doibase 10.21468/SciPostPhys.9.2.024} {\bibfield
  {journal} {\bibinfo  {journal} {SciPost Phys.}\ }\textbf {\bibinfo {volume}
  {9}},\ \bibinfo {pages} {024} (\bibinfo {year} {2020})}\BibitemShut {NoStop}%
\bibitem [{\citenamefont {Kuwahara}\ and\ \citenamefont
  {Saito}(2021)}]{PhysRevLett.127.070403}%
  \BibitemOpen
  \bibfield  {author} {\bibinfo {author} {\bibfnamefont {T.}~\bibnamefont
  {Kuwahara}}\ and\ \bibinfo {author} {\bibfnamefont {K.}~\bibnamefont
  {Saito}},\ }\href {\doibase 10.1103/PhysRevLett.127.070403} {\bibfield
  {journal} {\bibinfo  {journal} {Phys. Rev. Lett.}\ }\textbf {\bibinfo
  {volume} {127}},\ \bibinfo {pages} {070403} (\bibinfo {year}
  {2021})}\BibitemShut {NoStop}%
\bibitem [{\citenamefont {Green}\ \emph {et~al.}(2022)\citenamefont {Green},
  \citenamefont {Elben}, \citenamefont {Alderete}, \citenamefont {Joshi},
  \citenamefont {Nguyen}, \citenamefont {Zache}, \citenamefont {Zhu},
  \citenamefont {Sundar},\ and\ \citenamefont
  {Linke}}]{PhysRevLett.128.140601}%
  \BibitemOpen
  \bibfield  {author} {\bibinfo {author} {\bibfnamefont {A.~M.}\ \bibnamefont
  {Green}}, \bibinfo {author} {\bibfnamefont {A.}~\bibnamefont {Elben}},
  \bibinfo {author} {\bibfnamefont {C.~H.}\ \bibnamefont {Alderete}}, \bibinfo
  {author} {\bibfnamefont {L.~K.}\ \bibnamefont {Joshi}}, \bibinfo {author}
  {\bibfnamefont {N.~H.}\ \bibnamefont {Nguyen}}, \bibinfo {author}
  {\bibfnamefont {T.~V.}\ \bibnamefont {Zache}}, \bibinfo {author}
  {\bibfnamefont {Y.}~\bibnamefont {Zhu}}, \bibinfo {author} {\bibfnamefont
  {B.}~\bibnamefont {Sundar}}, \ and\ \bibinfo {author} {\bibfnamefont {N.~M.}\
  \bibnamefont {Linke}},\ }\href {\doibase 10.1103/PhysRevLett.128.140601}
  {\bibfield  {journal} {\bibinfo  {journal} {Phys. Rev. Lett.}\ }\textbf
  {\bibinfo {volume} {128}},\ \bibinfo {pages} {140601} (\bibinfo {year}
  {2022})}\BibitemShut {NoStop}%
\end{thebibliography}%

\end{document}